\documentstyle[epsf, twoside]{esapub}

\newcommand{\gw}{gravitational wave}
\newcommand{\grad}{gravitational radiation}

\newcommand{\fracparen}[2]{\left(\frac{#1}{#2}\right)}

\newcommand{\twopii}{2\pi\imath}

\newcommand{\E}[1]{\times 10^{#1}}

\newcommand{\splA}[2]{#2}
\newcommand{\splB}[2]{#1}
\newcommand{\American}{\let\spl=\splA} 
\newcommand{\British}{\let\spl=\splB} 

\newcommand{\Eqref}[1]{Equation~(\ref{#1})}

\newcommand{\Secref}[1]{Section~(\ref{#1})}
\newcommand{\hide}[1]{}\newcommand{\units}{\rm\;}

\newcommand{\Cardiff}{Department of Physics and Astronomy, University of Wales College of Cardiff, Cardiff, UK}
\newcommand{\AEI}{Max Planck Institute for Gravitational Physics (Albert Einstein Institute), Potsdam, Germany}

\newsavebox{\rubbish}

\newcommand{\e}{{\rm E}}

\begin{document}

\setlength{\parindent}{0pt}
\setlength{\parskip}{ 10pt plus 1pt minus 1pt}
\setlength{\hoffset}{-1.5truecm}
\setlength{\textwidth}{ 17.1truecm }
\setlength{\columnsep}{1truecm }
\setlength{\columnseprule}{0pt}
\setlength{\headheight}{12pt}
\setlength{\headsep}{20pt}
\pagestyle{esapubheadings}
\twocolumn
\title{\bf INTRODUCTION TO THE ANALYSIS OF LOW-FREQUENCY GRAVITATIONAL WAVE DATA\thanks{%
To be published in the Proceedings of the 1997 Alpbach Summer School on Fundamental 
Physics in Space, ed. A Wilson, ESA (1997).}}

\author{{\bf B.~F.~Schutz} \vspace{2mm} \\
{\bf \AEI} \and {\bf \Cardiff } }

\maketitle

\begin{abstract}

The space-based gravitational wave detector LISA will observe in the low-frequency 
gravitational-wave band (0.1~mHz up to 1~Hz).  LISA will search for a variety of expected signals, 
and when it detects a signal it will have to determine a number of parameters, 
such as the location of the source on the sky and the signal's polarisation.  
This requires pattern-matching, called matched filtering, 
which uses the best available theoretical predictions about the characteristics 
of waveforms. All the estimates of the sensitivity of LISA to various sources 
assume that the data analysis is done in the optimum way.  
Because these techniques are unfamiliar to many young physicists, I 
use the first part of this lecture to give a very basic introduction to time-series 
data analysis, including matched filtering.  The second part of the lecture applies 
these techniques to LISA, showing how estimates of LISA's sensitivity can be made,  
and briefly commenting on aspects of the signal-analysis problem that are special 
to LISA. \vspace {5pt} \\


  Key~words: gravitational waves, LISA, data analysis

\end{abstract}

\section{INTRODUCTION}

LISA, or any other space-based detector operating in the low-frequency regime, will take data continuously for a number of years (\cite{LISA1996}).  It will look simultaneously at the whole sky, with varying sensitivity in different directions.  It will record all the waves of sufficient strength in its observational frequency band that pass through the solar system during the lifetime of the mission.  But these waves will generally not be strong enough to be visible in the time-series data.  The output of the detector will simply look like pure noise most of the time.

To recognize and extract the signals, one must apply special computer operations, called filters, to the data to remove the noise and retain the signal.  Filters are constructed from theoretical expectations 
of what the waveform will  look like.  All our predictions of the performance and sensitivity of LISA 
assume that the data analysis systems implement filtering in an optimum way.  It is therefore 
impossible to understand how the LISA sensitivity is arrived at without knowing the elements 
of the theory of time-series data analysis, and especially of filtering.  

A filter that is known to most physicists, and which provides a simple example of the principle of 
filtering, is the Fourier transform.  If one expects the data to contain a simple signal of constant but unknown frequency, then the Fourier transform is the ideal filter to use to find it.   
Even if, in the time-series data, the 
signal amplitude is too small to be seen against noise, the Fourier transform allows one to identify the signal: it rearranges the data in such a way that the noise is spread over the whole spectrum but the power in the signal is concentrated at one frequency.   After applying this filter, the ratio of the signal's Fourier amplitude to the standard deviation of the noise at nearby frequencies can be very large.  Moreover, the filter has given us additional information about the signal: its amplitude, frequency, and phase, all of which can be regarded as parameters of the expected signal waveform that we wanted to determine.

Unfortunately for LISA, we do not expect its data to contain gravitational waveforms of constant frequency.  We do expect constant-frequency sources, such as binary systems, but the motion of the detector as it orbits the Sun imposes a Doppler shift on the incoming wave.  In the data itself, no physical signal will remain of constant frequency.  Moreover, many  other sources, such as 
the coalescences of supermassive black hole binaries, produce more complicated  signals whose frequencies depend on time in a predictable way.  Therefore, the filters that we will need for LISA must be more sophisticated than the simple Fourier transform.  Their construction and application is the subject of this lecture.

I will build on previous lectures at this school, which have covered the theory of general relativity and of gravitational waves, the design of the LISA detector, and the nature of the likely sources of low-frequency waves.  As in my previous lecture on sources, I will assume that many students have no previous 
expertise in this area, so in the first part of the lecture I will introduce the fundamentals of  
signal-analysis principles, including matched filtering.  In the second part I will describe how 
signal-analysis principles are used to estimate the sensitivity of LISA to various sources, and I will 
highlight some special signal-analysis problems that LISA will face.  

\section{ANALYSIS OF TIME-SERIES DATA}

\subsection{Detection is a Statistical Problem}

The data coming from LISA will be a single time-series of 
data values, sampled uniformly in time.  For the purposes of this lecture, I 
will assume that this data set has been calibrated and reduced to a data 
stream that directly corresponds to the amplitude $h$ of a gravitational wave that 
would be required to produce the observed strain $\delta\ell/\ell$ in the arms 
of either of the two LISA interferometers.  We do not need here to worry about 
instrumental issues, like the extraction of the data from the various signals 
within the detector, or the conversion of the data from measurements of changes 
of arm-length to measurements of gravitational wave amplitudes.

The  primary problem of the detection  of  \grad\ consists  of  identifying  a gravitational  
waveform  in  a  {\em noisy  signal}.  The noise can be intrinsic to the 
detectors itself  (vibrations or photon-counting 
fluctuations) or it can be external interference (a gravitational wave background due to 
cosmic sources or local binaries).  If nothing is known ahead of time about the signal, 
then the only way to detect it is to look in the time-series data for signals that are so strong 
that one would not expect noise to duplicate them during the observation period.  

It is possible to recognise much weaker signals if we know what 
to expect. However, even if  
the form of the wave is known, it may still depend on a number of 
parameters whose values in any particular event are unknown.  The amplitude and polarisation 
of the wave are always unknown parameters.  Other typical parameters include the time-of-arrival of 
a signal that has a limited duration, or the masses of the stars in a binary system that emits radiation. Since the waveform will be different for different parameters, finding the signal usually involves estimating its parameters at the same time.  This is where much of the scientific return of the LISA 
mission will be.

Because all data streams contain random noise, the detection of a signal is always 
a decision based on probabilities.  It is never possible to be 100\% sure of a detection: there  
is always a chance that the noise conspired in some random way to look very convincingly 
like an expected signal.  The aim of detection theory is to assess this probability, so 
that a confidence level can be attached to any claim of a detection.  

From the point of view of the data stream, there is nothing special about a signal: the data 
are just a series of random values.  The experimenter must decide what pattern in the data 
stream constitutes a 
signal and what does not.  Every signal should be defined before the analysis begins.  
In some cases, the definition is very precise, such as that the signal must look like 
that expected from a coalescing black-hole binary, within some range of parameters.  In 
other cases, the definition could be more inclusive, such as looking for a signal with 
an arbitrary waveform that is so strong in the time-series noise that one would not expect 
it to arise by chance more than once in, say, $10^4$~years of observing.  The more 
precise the definition of the signal, the less likely it is that random noise will duplicate 
it, and so the deeper into the noise one can go to find it.  
Experimenters should agree on the set of signal they are looking for before they 
analyse the data.

One analyst's signal could be another's noise.  A gravitational wave hunter would love 
to find the coalescing binary waveform, but would throw away features in the data 
that are clearly of instrumental origin.  The instrument-builder, on the other hand, might  
look precisely for such features as clues to the behaviour of the detector.  Even 
the noise could be a signal: we expect LISA to see a random background of waves 
from galactic binaries that is larger at some frequencies than the intrinsic detector noise.  

It is therefore important for the student approaching this subject for the first time to 
grasp the notion that signals are defined by the analyst, not by the experiment.  The 
analyst simply  looks for a pattern that seems close enough to his or her pre-conceived 
notion of a signal.  Any such identification has an element of chance in it, and one 
must assess one's confidence that the signal really did arise from an external source 
rather that something inside the detector.  The less information one uses in defining the 
expected signal (as when one is looking for new and unexpected sources), the easier 
it will be for noise to fit the definition, and the stronger must be the signal to be convincing.

\subsubsection{An example of signal identification pitfalls}

It may be helpful to illustrate the difference between an experiment that 
defines its signal before looking at the data and one that does not 
with a trivially simple problem.  It is often observed that any ``random'' 
sequence of numbers contains many patterns: what is different between 
random and deterministic numbers is that one cannot predict the patterns 
in the random number sequence.  To be concrete, suppose 
a student uses a computer program to generate a sequence of 10 integers, each one 
chosen randomly (and with uniform probability) from the integers 0..9.  The output 
of the first run of this program is the sequence (3, 7, 0, 1, 2, 3, 0, 9, 6, 1). The student  
notices right away that the sequence contains the subsequence 
(0, 1, 2, 3) in that order, beginning at the third integer in his list.  
He did not expect this.  Is it significant in some way?

He calculates that the probability that this sequence will appear, starting with the third 
element in the list, is $10^{-4}$.  But he reasons that this is not a fair assessment of 
the chances that the first sequence would surprise him: he would have been equally 
surprised had the sequence started at the second integer, or the fifth.  Since there are 
seven possible starting positions that would allow the whole sequence to appear, the 
the chance that this subsequence would appear somewhere in the data is 
0.0007: this will happen about once in every 1400 experiments if the random-number 
generator is working correctly.  This tempts the student to suspect an error 
in the program that generates the numbers.  But is this yet the true 
significance?  He reasons that he would also have noticed something unusual if 
the sequence had been (1, 2, 3, 4), or (9, 8, 7, 6).  When he calculates the 
chances of getting any such sequence in the data, then he gets about 0.01.  
He begins to think maybe he was just unlucky that this happened on the first 
run of his program.  But he really does not know how to assess the 
significance of this event.

In this case the solution to his decision problem is simple: he should run the 
program again, this time choosing to look for the ``signal'' (0, 1, 2, 3) 
somewhere in the ten integers produced.  Now the probability of obtaining 
this signal from a good random-number generator really is 0.0007. So if the 
signal appears again in the next sequence, it is very probable that the program 
is defective.  If 1400 different experiments like this were performed with 
the same result, then in 1399 of them the experimenter would be justified 
in throwing away the program and buying a new one.  Of course, it is still 
possible that our student will be the unlucky 1400th who arrived at this 
result at random!

What happens if something analogous occurs in the LISA analysis?  Particularly, 
what if one has found an unusual and unexpected ``signal'' using all 
the LISA data, so the option of going back and making another measurement 
is not there?  Or what if the source of the ``signal'' could be transient, 
so that further data taking would not be expected to see another such 
event anyway?  In order to minimise the chance that this will happen with 
LISA (or with other such experiments), it is important to define before the 
experiment what data sequences will be regarded as signals, so that their 
probabilities can be unambiguously calculated (like the number 0.0007 above).  
Even if the criterion is crude, to allow one to discover unexpected waveforms, 
one must define the criterion and calculate the chances of random noise 
satisfying it.  If this is not done, then one runs the risk of not being able 
to decide the significance of an observation.

\subsubsection{Literature}

There is a big literature on signal-analysis, much of it in electrical engineering.  
At least in part because this problem is very similar to  the problem 
of the detection of a target by radar  and  of  the estimation  of 
target parameters like range and velocity, the theory of detection and 
estimation is well understood.  Standard  textbooks  on  the  theory  
of  signal   detection include a number of monographs 
(\cite{Helstrom1968,VanTrees1968,Whalen1971,WainsteinZubakhov1962}).  
These texts  are  all oriented towards  applications  to radar.  The 
first introduction to signal-analysis theory aimed at the detection  of gravitational 
waves by broadband detectors is a review article by \cite{Davis1989}. This article has 
the additional advantage of being written in  the  contemporary 
language of stochastic processes.  The techniques of time-series 
analysis (Fourier transforms, etc.) are fundamental tools of the trade.  
There are many introductory textbooks, such as 
\cite{bracewell}.  I have previously reviewed 
these issues principally in the context of ground-based interferometers 
(\cite{schutzblair,schutzleshouches}).

\subsection{Signals in Noise}

As a result of noise a datum from the detector is  a  value  of  a 
certain {\em random variable}.    Since we take measurements at 
regular intervals of time, the data from a detector form a sample of a 
certain (discrete)  {\em stochastic process}.  Regardless of whether a 
signal is present or not, the data will still be stochastic, but the  
presence  of a signal will affect the probability  distribution  
of  the  stochastic process.  

We will use the following notation: $x_j$ denotes a sample of the 
experimental data (the random variable),  and 
${\bf x} = \left(x_0, x_1, ..., x_{N-1}\right)$ denotes the entire data set of $N$ samples, which 
is called the stochastic  {\em process}.  The expected signal will be 
denoted by $h_j$ and the noise (which would be the output in the 
absence of a signal) is called $n_j$.  Sums, where no limits are 
indicated, always run from $0$ to $N-1$.

\subsubsection{Detection}

In general, the distribution of data values $x_j$ will be described by some probability 
density function (abbreviated pdf) $p(x)$: 
the probability that $x_j$ lies between a value $x$ and a value $x+dx$ 
is $p(x)dx$.  The probability for the whole set of values (process) is 
the joint pdf of the process.  If the points are statistically independent, 
the joint probability is the product of the individual probabilities.  But 
data values are not always (in fact, not usually) statistically independent.  

Since the output of the detector depends on whether a signal is present or 
not, the pdf of the data must be different in the two cases.  
If  there  is  no   signal we   call the joint pdf $p_0({\bf x})$; if the signal is  
present we call it $p_1({\bf x})$.   We will 
see below an example of how to calculate these pdf's.  
To decide which pdf applies to a 
particular measurement $\bf x$, we have to devise a rule called the 
{\em test}, which decides whether the observed data were more likely to come from 
a distribution with pdf $p_0$ or $p_1$. It does this by dividing 
the range of possible values of ${\bf x}$ into two sets 
$R$ and its complement $R'$ in such a way that we decide the pdf is 
$p_1({\bf x})$ if ${\bf x} \in R$ and the pdf is  $p_0({\bf x})$ if 
${\bf x} \in R'$.

The {\em detection probability} $P_D(R)$ is then given by the 
probability that a data set $\bf x$ that contains the signal will pass 
our test: 
\begin{equation}\label{eqn:detprob} 
P_D(R) = \int_{R} p_1({\bf x}) \,d{\bf x} .
\end{equation} 
The {\em false alarm probability} $P_F(R)$ is the  probability that a 
data set that contains no signal passes our test:
\begin{equation} \label{eqn:faprob} 
P_F(R) = \int_{R} p_0({\bf x})\,d{\bf x}. 
\end{equation}

\subsubsection{Confidence and significance of a detection: the null hypothesis}

As we have said above, detection is always probabilistic: one can never be 100\% certain 
that a signal is present.  There is always the possibility that the noise 
has conspired to look like the desired signal.  The only legitimate 
statement of the outcome of an experiment is to give a probability for 
the signal to be present.  

What is often done is to quote the false-alarm probability, which is the probability  that 
the experiment would have had the same outcome even if no signal had been 
present.  This is sometimes called the null hypothesis.  Such calculations are 
delicate, since the probability of getting a false alarm depends on how many times 
the experiment is performed.  When one is looking for a family of signals that 
depends on a parameter, such as the masses of the stars in a binary system, 
then each search through the same data set for each distinguishable value of the parameter is a different 
experiment, and the probability of getting a false alarm increases directly with the 
number of parameter values.

Sometimes the number of experiments is not easy to assess.  When physicists 
look through data and find unexpected results, which are not part of a 
previously defined family of signals, it may be impossible to decide just how big 
the parameter space was, how big the false-alarm probability was. We 
illustrated this point in the introduction above.  For a real-life 
discussion of these issues in the context of gravitational wave detection, see \cite{dickson} and 
references therein.  For this reason, it is important for experimenters to define what will 
be accepted as a signal before they look at the data.

The false alarm probability  is called the  
{\em significance}.  The {\em confidence level} is one minus this.  
It is clear that, even in cases where the number of 
experiments is well-defined, the calculation of the false-alarm probability 
requires a full understanding of the properties of the noise in the experiment. 
{\em Characterising the noise is a vitally important part of any experiment.} 

\subsubsection{Systematic errors}

Until now we have assumed that the data stream consists of random noise 
plus a possible signal.  But it could be influenced by other effects too.  
For example, it could contain a coherent ``signal'' that looks like an 
expected astronomical gravitational wave but really comes from 
an instrumental source.  Or the 
external disturbances that set the low-frequency limit on  the LISA 
band might contain occasional large events associated with 
fluctuations in the solar wind or radiation pressure.  This is an 
unavoidable problem in data analysis, and there are no general prescriptions 
for dealing with it.  One must build in as many housekeeping data streams and 
consistency checks as possible, to give confidence that systematics play 
no important role.  Again, understanding all the details of the instrumental 
noise will be essential to the success of LISA.

\subsubsection{Neyman-Pearson test: the likelihood ratio}

The most appropriate way to test for the detection of  gravitational 
waves seems to be the {\em Neyman-Pearson} approach.  In this approach we 
seek a test that maximises the detection probability subject to a 
preassigned false alarm probability $P_F(R) = \alpha$.  Given that \gw 
s have not yet been detected, an important consideration at least at 
first will be to be sure that one has detected one.  By choosing the 
false-alarm probability $\alpha$ sufficiently small, one ensures  that 
the chances of falsely identifying a noise event as a gravitational 
wave are as small as one wants.  In this way one misses many true 
gravitational wave events, but one has considerable confidence that 
those that are identified are real.  

The solution for the ``detection region'' $R$ in the Neyman-Pearson 
approach is given in terms of the {\em likelihood ratio}, defined by
\begin{equation} \label{eqn:likelihood}
\Lambda({\bf x}) = \frac{p_1({\bf x})}{p_0({\bf x})}. 
\end{equation} 
If the observed data set ${\bf x}$ is more easily produced when a 
signal is present than when absent, then this will be larger than 1.  
The larger this is, the more likely it is that a signal is present.  
If we let $\Lambda_0$ be the {\em likelihood threshold} associated with the 
probability $\alpha$, 
\begin{equation} \label{eqn:kdef} 
P_F[\Lambda({\bf x}) \geq \Lambda_0]  = \alpha, 
\end{equation} 
then the detection region of the space of all possible sample sets is 
\begin{equation} 
R = \{{\bf x} : \Lambda({\bf x}) \geq \Lambda_0 \}. 
\end{equation} 

Thus if for a particular observed set $\bf x$ we find that 
$\Lambda({\bf x})\geq \Lambda_0$, then we say that the signal is present; 
otherwise we  say  that  the signal is absent. 

In most of the  following we assume that the sampled values of the signal 
$h_i=h(t_i)$ are a deterministic function of 
time that we expect to find in our data, for example because 
calculations of gravitational wave sources have revealed it. We will 
have to treat separately the case where the ``signal'' itself is noise, 
such as a background generated by the Big Bang.

\subsubsection{Characterisation of noise}

We assume now that the noise $n_j = n(t_j)$  in  the  detector  is  a  zero-mean
Gaussian stochastic process.   In fact, real noise is not completely Gaussian, 
and this needs to be taken into account when systems are designed for LISA.  
But for this introductory lecture, it will be best to assume the simplest 
model of noise. The main way of 
characterising this noise is by its {\em autocorrelation function} 
$K_{ij}$: 
\begin{equation}\label{eqn:corrfcn} 
K_{ij}=\e[n_in_j], 
\end{equation} 
where $\e[ \;]$ denotes the expectation value. If the noise is zero-mean Gaussian, then the 
autocorrelation function completely determines the statistics of the process. 

\subsubsection{Matched filtering of Gaussian noise}

It can be shown for 
such a process that the  logarithm  of  the likelihood ratio is given 
by 
\begin{equation} 
\ln \Lambda[{\bf x}]  =\sum_{k}x_kq_k  - 
\frac{1}{2}\sum_{k}h_kq_k,\label{eqn:LR} 
\end{equation} 
where $q$ is called the {\em matched filter} for the expected signal $h$. 
It is defined to be the solution of the equation 
\begin{equation} 
h_j = \sum_{k}K_{jk}q_k. \label{eqn:int1} 
\end{equation}
This is a matrix inversion for $q$, but if there are many data points the 
inversion may not be easy to perform.  
We will see below how to solve this equation for $q$ more easily in 
the case of stationary noise. The waveform $h$ that is used to generate $q$ is called 
the {\em template} for $q$. 

Notice 
that the solution for $q$ does not depend on the observed data set $x$, but 
only on the statistical properties of the noise as given 
by $K_{ij}$.  Once one has 
solved for $q$, one then computes the logarithm of the likelihood
ratio from \Eqref{eqn:LR} for any particular data set.  If this 
exceeds the threshold $\ln\Lambda_0$, one claims a detection.

In fact, we can see that the likelihood ratio $\Lambda[{\bf x}]$  depends on the 
particular set of data ${\bf x}$ only through the sum
\begin{equation}
G = \sum_{k} x_kq_k. \label{eqn:Gdef}
\end{equation}
This sum $G$ is called the {\em detection statistic} for the signal $h(t_j)$ which 
generates the filter $q(t_j)$ through \Eqref{eqn:int1}.  
The second term in \Eqref{eqn:LR} is independent of the data set, so we do not 
include it in the detection statistic.  The statistic is a correlation between the data $x$ 
and the filter $q$.  

Since we want to maximise 
the likelihood ratio, we want to maximise the detection statistic: if $G$ is larger than 
one expects by chance, then there is a corresponding likelihood that the data contains 
the signal $h$.  It is important, therefore, to note that the detection statistic is a 
dimensionless number: if $x$ has units km, say, then the autocorrelation 
$K$ has units $\units km^2$, the filter $q$ defined in \Eqref{eqn:int1} 
has units $\units km^{-1}$, and the statistic $G$ is dimensionless.  
We will see later that significant detections are associated with 
observed values of $G$ that are large compared to $1$.

\subsection{Data Analysis in the Presence of Stationary Noise}

A very important special case is that of {\em stationary} noise, which 
is defined as a process which is independent of the origin of time, 
{\em i.e.} of when the experiment started.  The autocorrelation 
then depends only on the {\em difference} between the times $t$ and 
$t'$: there exists a $C_k$ such that 
\begin{equation}\label{eqn:corrfcnstat}
K_{ij} = C_{i-j}. 
\end{equation}
LISA will normally observe long-lived sources, so the assumption 
of stationarity is a strong one: most instruments change their sensitivity 
over time.  Again, for the  purpose of this lecture, this assumption is a good 
first step.  But a full data analysis program for LISA must address 
the best way to achieve long-term sensitivity as the detector changes.

The great advantage of stationary noise is that, if the noise is 
stationary and  if  the whole of the signal is included in the interval  
$[0,T]$,   then  \Eqref{eqn:int1}  can be solved explicitly for the 
filter by Fourier transform techniques.  To see how this happens 
we must first review some part of the theory of discrete Fourier transforms.

\subsubsection{Definition of the discrete Fourier transform}

Given any data series $x_j$ of $N$ points numbered from 0 to 
$N-1$, we define its discrete Fourier transform (DFT) $\tilde{x}_k$ 
by 
\begin{equation}\label{eqn:FT}
\tilde{x}_k=\sum_{j=0}^{N-1}x_je^{-\twopii jk/N}.
\end{equation}
Its inverse is
\begin{equation}\label{eqn:FTinv}
x_j=\frac{1}{N}\sum_{k=0}^{N-1}\tilde{x}_ke^{\twopii jk/N}.
\end{equation}
The factor of $1/N$ must not be forgotten.  There are various conventions 
for these definitions, according to whether $1/N$ is placed in the Fourier 
transform or in its inverse, or is shared between them.

If the index $j$ counts sampled points in the time domain, the index $k$ is 
the frequency index.  Since a given real signal has Fourier components 
at both positive and negative frequencies (as in $\cos(\twopii ft) = 
\frac{1}{2}(\exp(\twopii ft)+\exp(-\twopii ft)$, which has components at frequencies $\pm f$), 
both positive and negative frequencies are present in the Fourier transform, 
with complex-conjugate amplitudes.  With our conventions, 
the indices $k=0$ to $k=N/2$ denote the positive frequency components. 
The negative frequencies are mapped by periodicity to larger values of $k$: the negative 
component associated with a positive frequency $k$ is at $N-k$.  

\subsubsection{Relation to the continuous Fourier transform}

Physically, one usually likes to think in terms of the continuous Fourier 
transform of a data stream $x(t)$ that our discrete values $x_j$ were sampled from: 
\begin{equation}\label{eqn:continFT}
\tilde{x}( f ) = \int_{-\infty}^{\infty}x(t)e^{-\twopii  f  t}dt.
\end{equation}
I will use a notation from now on that the tilde $\tilde{}$ denotes a Fourier transform, 
and the type of transform is given by the argument: if the argument is the 
continuous variable $ f $ it is the continuous transform just defined, and if 
the argument is a discrete index like $k$ it is the DFT.

Here we have continuous variables for time $t$ and frequency $f$.  The first correspondence we 
must make is between them and their discrete counterparts 
$j$ and $k$.  We shall always assume that the time-series data were sampled at a uniform rate, 
with a sampling interval $\Delta t$.  Then we have $t_j = j\Delta t$.  (For convenience, we 
take the origin of time $t=0$ to be at the first sampled data point.)  The duration 
of the experiment is $T= N\Delta t$.  The frequency associated with a given Fourier 
amplitude $\tilde{x}_k$ must be deduced by equating the argument of the exponential 
in the definition of the DFT  
in \Eqref{eqn:FTinv}, $\twopii jk/N$, to that in the continuous Fourier 
transform in \Eqref{eqn:continFT}, $\twopii  f  t$.  Replacing $j$ by its associated time 
$t_j/\Delta T$, we find that  
\begin{equation}\label{eqn:freqdef}
 f_k = k/(N\Delta t) = k/T.
\end{equation}
Therefore, the frequency resolution of the data is $\Delta  f  = 1/T$.   The DFT essentially 
groups the Fourier components of the data into discrete Fourier bins of width $1/T$.

The second correspondence is to work out the relationship between values of the continuous 
Fourier transform and of the discrete version.  The definition of the DFT does not contain an integration 
element corresponding to $dt$ in the continuous FT, using instead a simple dimensionless summation. 
This integration interval has width $\Delta t$, so we have
\begin{equation}\label{eqn:DFTandcont}
\tilde{x}( f_k) \rightarrow \Delta t \tilde{x}_k.
\end{equation}
The arrow reminds us that this is not an equality, but a correspondence.  The full continuous Fourier 
transform uses the whole, infinitely long data set, while the DFT can only approximate it from a  
finite set of sampled values.

\subsubsection{Properties of the DFT}

The DFT is nothing more than a complex Fourier series representation of the 
finite set of sampled data values.  As a Fourier series, it assumes periodicity: 
if one calculates the values of $x_j$ for $j>N$ from the inverse series \Eqref{eqn:FTinv}, 
they will simply continue the sampled data forward in a 
periodic way.  If the underlying continuous data contains a signal whose spectrum 
as defined by \Eqref{eqn:continFT} has a frequency $f_0$ that is 
exactly equal to one of the resolved frequencies in the DFT, say $ f_0 = k_0/T$, 
then the DFT will have two non-zero elements at $k_0$ and $N-k_0$.  If 
the true signal frequency $f_0$ falls between two discrete frequencies, then the spectrum 
will be more complicated, but it will be dominated by the amplitudes at the 
two nearest discrete frequencies. 

The DFT is a linear operation, so in particular when applied to noisy data it commutes with the expectation operation:
\[\e\left[\tilde{x}_k\right] = \sum_{j=0}^{N-1}\e\left[x_j\right]e^{-\twopii jk/N}.\]
There are a number of other important results that are easy to derive.  One that we will need later 
is that if the data set is run backwards in time, the DFT is just the complex conjugate of the 
original.

\subsubsection{Convolutions using the DFT}
One of the most important operations one can perform with Fourier transforms is 
the {\em convolution}.   Note  
that with a data set that is finite, the only way to define a convolution is cyclically: 
the data wrap around from back to front as they are shifted, or equivalently 
they extend to indices outside the range $(0,\ldots,N-1)$ periodically.  Thus, the convolution of 
two discrete data sets $\{h_j,\,j=0,\ldots,N-1\}$ and $\{g_{j},\,j=0,\ldots,N-1\}$ is defined as 
\begin{equation}\label{eqn:discreteconvdef}
({h}\circ{g})_j = \sum_{j'=0}^{N-1}h_{j'}g_{j-j'}=({g}\circ{h})_j.
\end{equation}
The key result here is the {\em convolution theorem}: 
\begin{equation}
\left(\widetilde{h\circ g}\right)_k =\tilde{h}_k\tilde{g}_k. \label{eqn:discreteconvthm1}
\end{equation}
This allows us to perform convolutions by doing Fourier transforms.  

In the convolution, the index $j'$ runs in opposite directions through the sets $h$ and $g$.  
A correlation is defined as the same sum with the index on $g$ reversed:
\begin{equation}\label{eqn:discretecorrdef}
\mbox{corr}(h,g)_j = \sum_{j'=0}^{N-1}h_{j'}g_{j'-j} = \mbox{corr}(g,h)_{N-j}, 
\end{equation}
where again indices are extended outside the range $(0,\ldots,N-1)$ by making the functions 
periodic.  The equivalent of the convolution theorem is now 
\begin{equation}
\left(\widetilde{\mbox{corr}(h,g)}\right)_k =\tilde{h}_k\tilde{g}_k^*, \label{eqn:discretecorrthm1}
\end{equation}
where the ${}^*$ denotes complex conjugation, which (as remarked above) changes the 
transform of $g_j$ into that of $g_{N-j}$.

\subsubsection{Power spectrum and power spectral density}

To work with noise in the Fourier domain, we need to characterise its probability 
distribution at any frequency, that is we need to understand the spectral noise $\tilde{n}_k$, 
the transform of the time-series noise $n_j$.  Of course, for zero-mean noise the expectation 
of the spectral noise amplitude will vanish.  But its square will not, and the 
expectation of the squared modulus of the noise is (to within a factor of 
2) the variance of the spectral noise distribution.  

Now, the power spectrum of a data set is defined as the 
squared magnitude of the Fourier transform:
\[X_k = \left|\tilde{x}_k\right|^2.\]
Many authors call this the periodogram, but we will reserve that term for 
something a little different (below).  

When the data is noisy, then we are interested in expectations of the power.
For completeness, we consider the expectation of the product of the Fourier 
amplitudes of the noise alone at two (possibly) different frequencies:
\[P_{kk'} = \e \left[\tilde{n}_k\tilde{n}^*_{k'}\right] = \sum_{j=0}^{N-1}\sum_{j'=0}^{N-1}\e\left[n_jn_{j'}\right]e^{-\twopii (jk-j'k')/N}.\]

If the noise is {\em stationary}, one can use \Eqref{eqn:corrfcn} and \Eqref{eqn:corrfcnstat} to obtain 
\[P_{kk'} = \sum_{j,j'}C_{j-j'}e^{-\twopii (jk-j'k')/N}.\]
By changing variables so that $j-j'$ is a summation variable and then doing the other summation explicitly\footnote{One must use the identity $\sum_k \exp(-\twopii jk/N) = N\delta_{j0}$, which is proved 
using the sum formula for a geometric series.}, one can show that 
\begin{equation}\label{eqn:psd}
\e \left[\tilde{n}_k\tilde{n}^*_{k'}\right]  = S_k\delta_{k'k}, 
\label{eqn:discretepdsdef2}
\end{equation}
where the {\em power spectral density} (psd) $S_k$ is given in terms of the 
time-autocorrelation $C_j$ by 
\begin{equation}\label{eqn:psddef}
S_k = P_{kk} \;\mbox{(no sum on $k$)}\; = N\sum_{j=0}^{N-1}C_je^{-\twopii jk/N}.
\end{equation}
The psd can be estimated from observed data by constructing the power spectrum 
and using any of a variety of methods to estimate from this the expectation value in \Eqref{eqn:psd}.  
One method is to average several successive data sets; another is to smooth the power 
spectrum by averaging groups of frequencies.  In each case, one must convince oneself 
that the data are not ``contaminated'' by signals, that the expectation is really the expectation 
of the noise.

\Eqref{eqn:psd} shows that, for stationary noise, there is no correlation between the noise at 
different frequencies, and at a given frequency the psd is the expectation of the ordinary power 
spectrum.     We say that the noise is {\em white} if the psd is flat, so that 
there is the same power at all frequencies.  In this case, the inverse transform of 
\Eqref{eqn:psddef} shows that $C_j = 0$ for $j\neq 0$.  This means that, for white 
noise, data samples in the time-series are not correlated with one another.  Many sources of noise are 
intrinsically white, but most measuring instruments have different responses 
at different frequencies, so that the noise in their outputs will generally be 
coloured.

\subsubsection{The matched filter for stationary noise}
Putting all these results together, we find that --- by the convolution theorem --- 
the solution $q$ of \Eqref{eqn:int1} has the Fourier transform
\begin{equation}\label{eqn:filtersolve}
\tilde{q}_k=N\frac{\tilde{h}_k}{S_k}.
\end{equation}
This filter is called the {\em matched filter} for the expected signal
$h$, and this equation is one of the most important in practical terms 
in signal-analysis theory.  Notice that it involves an inverse weighting with $S_k$: if the noise is
high at some frequency, the Fourier component of the filter at that
frequency will be reduced, so that it contributes relatively less to
the convolution that produces the statistic $G$ in \Eqref{eqn:Gdef}
that searches for the signal.  This shows how the detector's noise
distribution limits its sensitivity. 

The detection statistic is linear in the filter $g$, so filters are essentially 
equivalent if they differ by an overall constant: $g_j$ and $\beta g_j$ will 
find the same signals.  Of course, one has to take the filter normalisation into account 
when inferring the amplitude of the signal, which we address below.  Because 
of this, we will keep the normalisation indicated in \Eqref{eqn:filtersolve}, with 
its factor of $N$.  

Notice that, if the noise is white, then the matched filter is just the expected signal.  
It is therefore helpful conceptually to assume white noise when trying to 
understand what happens with filtering, even though in  a real case one 
can usually not make this assumption.

It should be noted that there are slightly different conventions for what 
one calls the matched filter.  Some authors call $q_k^*$ (the complex 
conjugate) the filter, or its transform.  Note that the time-domain version 
of this is essentially the signal running backwards.  This difference is 
just a matter of nomenclature: everyone will perform the same mathematical 
operations in order to filter for the signal.

\subsubsection{Parseval theorem}  \label{sec:parseval}

There is an important relation between the Fourier transform and 
the original time-series data.  This is called the Parseval theorem:
\begin{equation}\label{eqn:parseval}
\sum_{k=0}^{N-1}\left|\tilde{x}_k\right|^2 = N\sum_{j=0}^{N-1}\left|x_j\right|^2.
\end{equation}
By letting $x_j$ be the sum of two variables $y_j+z_j$, it is easy to 
deduce the more general form of this result:
\begin{equation}\label{eqn:parseval2}
\sum_{k=0}^{N-1}\tilde{y}_k\tilde{z}^*_k = N\sum_{j=0}^{N-1}y_jz_j.
\end{equation}

An interesting deduction from this is what happens to the DFT 
of a long-lived signal when the observation time $T$ increases.  The 
right-hand-side of this equation increases roughly as $N^2$, so 
that means that the total power in the signal (the left-hand-side) 
also increases with $N^2$ (or $T^2$). {\em If the signal has only one 
significant Fourier amplitude (a narrow-band signal), then 
this amplitude increases linearly with time.  }

Similarly, what happens to the noise psd $S_k$ when the observation 
time increases?  If we take the expectation of the previous equation 
with $x$ replaced by the pure noise amplitude $n$, then 
the left-hand side simply sums over $S_k$ and the right-hand side 
adds up the expectation of the squares of the random values $n_j$.  On 
the right, for stationary  noise, the expectation for each $j$ is the same, 
so the sum is proportional to the number of data samples $N$.  In addition, 
there is the explicit factor $N$ on the right-hand-side, so the whole expression 
is proportional to $N^2$.  On the left-hand-side, for white noise (just to make 
the argument simple) all the values $S_k$ are the same, and there are $N$ of 
them.  There are no extra factors of $N$ there, so the expression is just 
$N$ times the typical value of $S_k$ for any $k$.  By equating this to the 
right-hand-side, we find that {\em the power spectral density $S_k$ itself is proportional to $N$, or to the 
duration of the experiment $T$.}

\subsubsection{Nyquist theorem and aliasing}

Notice that, for data sampled with an interval $\Delta t$ and therefore a 
sampling rate $1/\Delta t$, the range of frequencies extends only up 
to $k = N/2$, or $ f_k = 1/2\Delta t$.  So the maximum 
frequency that can be represented by data sampled at a given 
rate is half the sampling rate.  Conversely, if we want to represent 
the spectral content of a signal up to a frequency $ f_0$, we must 
sample the data at least twice as fast.  

The full statement of this relationship is Nyquist's theorem, which is 
that if an infinitely long continuous data stream contains no spectral power at 
frequencies above $ f_0$, then sampling that stream at a rate 
of $2 f_0$ is sufficient to capture all its information: the whole 
continuous stream, even values between sampled points, can be 
reconstructed from the sampled ones by using the inverse Fourier 
transform of the sampled points to generate time-series values at 
any value of the time, not just at a sampled time.  Then if the 
original signal was band-limited in its spectral content as assumed, 
the reconstructed values will be exactly what the original continuous 
stream would have given had it been sampled there.

This theorem is not very useful as it stands, because it assumes an 
infinitely long data set ($N=\infty$).  Every real experiment has a 
beginning and an end, and it generates a finite amount of data.  If 
the data stream is long compared to the signal duration, however, 
then one can expect the Nyquist theorem to apply.

How does one know that the data are band-limited in the first place?  The answer 
is that the experimenter usually arranges this by filtering out all frequencies 
above a certain value before the data are sampled.  This is to avoid 
a phenomenon called {\em aliasing}.  We shall define aliasing, 
and then see why it is something to avoid.

Consider what happens to 
the DFT in \Eqref{eqn:FT} if the data contains a signal $h$ that has a frequency that is higher 
than the highest frequency in the DFT, $N/2T$.  Suppose, for 
concreteness, its frequency is $(N/2 + k_0)/T$.  Then at $t_j = j\Delta t$ 
its value would be 
\[h_j = \exp\left(\twopii  f  t_j\right) = \exp\left(\twopii (N/2 + k_0)j/N\right).\]
By replacing $N/2$ by $N-N/2$ and using the fact that $\exp(\twopii j) = 1$, 
we find
\[h_j = \exp\left(\twopii (k_0 - N/2)j/N\right).\]
In other words, the signal at the high frequency of $k_0+N/2$ has the identical values 
at the sampling times as one at the lower frequency of $k_0-N/2$.  If the first frequency was 
outside the frequency band represented by our data, the second one could be inside it.  In this 
case, the high-frequency signal would be mapped onto one in our band, and we would not know 
that it was ``really'' a higher frequency.  If the second frequency is still outside our 
band, then we can do the calculation again, and map it to $k_0-3N/2$, and so on 
until it appears in our band.  It follows that a signal of {\em any} frequency, no matter 
how high, will be aliased by sampling into a signal with an apparent frequency in 
the bandwidth of our DFT.

One can think of aliasing in terms of periodicity.  The inverse Fourier transform, 
\Eqref{eqn:FTinv}, shows that the data values are periodic functions of time, since 
the time index $j$ enters the summation only in sinusoids that are periodic in $j$ with  
a period $N$.  In just the same way, the Fourier transform itself, \Eqref{eqn:FT}, 
shows that the Fourier coefficients are periodic functions of frequency, since 
the frequency index $k$ enters the summation only in sinusoids that are periodic 
in $k$ with a period $N$.  Aliasing says that, whatever the true spectrum of the continuous data 
is, the act of sampling it will add together frequency components from all the periods 
of the fundamental frequency band ($-1/2\Delta t$, $+1/2\Delta t$).  

This is bad in an experiment because, even if there are no interesting signals at 
higher frequencies, there is usually noise.  If the data are sampled without filtering 
this high-frequency noise away, it will contribute to the noise in the observation band 
by aliasing.  So experimenters generally filter out this noise, ensuring that the signal 
that is sampled already has zero (or minimal) power at frequencies above half of 
the sampling frequency.  This filtering is usually done by analogue means, using 
filter circuits in the electronics, because once the data have been sampled it is too late 
to get rid of the high-frequency components.

\subsubsection{Interpolation in time and in frequency: the periodogram}

Once we have band-limited, sampled data, we can apply the Nyquist theorem (at 
least approximately, since our data set is finite in duration) and try to 
reconstruct the data values between the sampled values.  
This {\em interpolation} is done just by 
using the inverse Fourier transform at intermediate values:
\begin{equation}\label{eqn:interpolateintime}
x(\alpha) = \frac{1}{N}\sum_{k=0}^{N-1}\tilde{x}_ke^{\twopii \alpha k/N}, 
\end{equation}
where $\alpha$ is a continuous variable related to the interpolation time $t$ by $t=\alpha\Delta t$.
Naturally, since the interpolation theorem is only strictly valid if the data set is 
infinite, when we use it on our finite set we must not use it outside the original 
duration of the experiment: the formula is for interpolation, not extrapolation!

Because of the deep symmetry between the Fourier transform and its inverse, every 
theorem about functions of time and frequency has a counterpart with time replaced 
by frequency and frequency by time.  We illustrated this in our explanation of aliasing 
in terms of periodicity in the frequency domain.  The interpolation result we have 
just quoted also has a useful counterpart that enables us to look for signals of 
any frequency, not just the discrete ones of the DFT.  It goes like this. 

Suppose we have a continuous data stream that is time-limited: the data stream really is zero before and 
after the time of the experiment.  Then the analogue of the Nyquist theorem is that 
it can be described completely by a discrete but infinite set of Fourier coefficients.  We can 
use the data values to interpolate between the discrete frequencies using the 
Fourier transform with a continuously variable frequency index.  Of course, this 
would be exact only if we used the infinite set of frequencies, and we have from 
our analysis only a finite set.  But if the data does not contain much information 
at higher frequencies (because of our anti-aliasing filter) then we won't be too far 
wrong to use
\begin{equation}\label{eqn:interpolateinfrequency}
\tilde{x}(\beta) = \sum_{j=0}^{N-1}x_je^{-\twopii j\beta/N},
\end{equation}
where $\beta$ is a continuous variable related to the interpolation frequency $ f $ by $ f =\beta T$.
This, or more commonly its associated power spectrum $|\tilde{x}_\beta|^2$, 
is called the {\em periodogram} of the 
data.  It can be used to look for narrow-band signals at frequencies between the discrete 
frequencies of the DFT.

The periodogram defined here is a function of a continuous variable $\beta$.  There is no easy way 
to compute it except by direct summation if a value is desired for some arbitrary value 
of $\beta$.  But if one wants a regularly sampled periodogram, {\em i.e.}  a refinement of 
the DFT by interpolating $m$ equally spaced points between the existing points of the 
DFT, then there is an efficient way.  This is to make an FFT of a data set that consists 
of the original sampled points extended (padded) by $mN$ zeros.  By analogy, this 
trick also works for the time-interpolation formula \Eqref{eqn:interpolateintime}.

It should be noted that there is no uniformity in the use of the term {\em periodogram} in the literature. 
Often it is used synonymously with what we have called the power spectrum, 
which means it is just evaluated at the discrete resolved frequencies of the DFT.  This 
is just a subset of the values of the periodogram as we have described it, but it is important 
to understand the nomenclature used by whatever author you are consulting.

\subsection{The Signal-to-Noise Ratio in Stationary Gaussian Noise}

\subsubsection{Probability distribution functions}

When the noise is Gaussian, which is a good enough starting assumption for us, 
we can easily compute $p_0$ and $p_1$ for the test based on computing 
the detection statistic $G$, given in \Eqref{eqn:Gdef}.   As a sum of Gaussian-distributed 
random numbers, $G$ itself is Gaussian.  We need only compute its mean and 
variance to determine the distribution.  Since we assume stationary noise, we 
take the Fourier form of the expression for $G$:
\begin{equation}\label{eqn:fourierG}
G = \sum_jx_jq_j = \frac{1}{N}\sum_k\tilde{x}_k\tilde{q}^*_k = \sum_k\tilde{x}_k\tilde{h}^*_k/S_k,
\end{equation}
where the second equality follows from the Parseval theorem \Eqref{eqn:parseval2}.  

If there is no signal, the data are pure noise ($x_j = n_j$), and so the mean of $G$ is 
\[\e[G] = \sum_k\e[\tilde{n}_k]\tilde{h}^*_k/S_k = 0,\]
for zero-mean noise.  For the variance then we have
\[\e[G^2] = \sum_k\sum_{k'}\e[\tilde{n}_k\tilde{n}^*_{k'}]\tilde{h}^*_k\tilde{h}_{k'}/(S_kS_{k'}).\]
Using \Eqref{eqn:psd} for the expectation value here, we define the variance by the symbol $d_0^2$:
\begin{equation}\label{eqn:filtervariance}
d_0^2 = \e[G^2]\mbox{ (no signal) } = \sum_{k=0}^{N-1}\frac{|\tilde{h_k}|^{2}}{S_k}.
\end{equation}
The pdf of $G$ in the absence of a signal is the zero-mean Gaussian with 
this variance:
\begin{eqnarray}
\mbox{absent:}\quad p_0(G) & = & (2\pi d_0^{2})^{-1/2}\;\times \nonumber \\
&&\exp(-G^2/2d_0^2) \label{eqn:statnone} 
\end{eqnarray}

If the signal $h$ is present, then the data consists of the signal plus the same 
type of noise as before.  The expectation  of the data is just the signal in the filter, 
\[\e[G] = \sum_k\left|\tilde{h}_k\right|^2/S_k = d_0^2,\],
and the variance is the same as before, since the signal is deterministic.  So the 
pdf of $G$ in the presence of the signal is
\begin{eqnarray}
\mbox{present:}\quad p_1(G) & = & (2\pi 
d_0^{2})^{-1/2}\;\times \nonumber \\
&&\exp[-(G-d_0^{2})^2/2d_0^{2}]. \label{eqn:statsig} 
\end{eqnarray} 

Notice that, as we observed before, $G$ is dimensionless; therefore so is $d_0$.  
The expected value of $G$ when a signal is present is $d_0^2$, while the standard 
deviation away from that value is $d_0$.  Therefore, the {\em signal-to-noise ratio} (snr) of the 
filter output is $d_0$.  Since $G$ is linear in the data $x$, this is the {\em amplitude} signal-to-noise 
ratio of the filtered data.  The {\em power} snr is the square of this:
\begin{equation}\label{eqn:snrsta}
(\mbox{amplitude snr})^2 = \mbox{ power snr } =  d_0^2 =   \sum_{k=0}^{N-1}\frac{|\tilde{h_k}|^{2}}{S_k}.
\end{equation}

It is important to recall here that we have not allowed the expected signal $h$ to depend 
on any parameters yet: even its amplitude is assumed fixed.  The detection problem assumes 
that either the signal is present with exactly the assumed amplitude, or it is absent and 
the data are pure Gaussian noise. The filter $q$ is constructed from a signal with 
the expected amplitude; and the snr of the filter output, $d_0$, also depends on the amplitude of the 
expected signal $h$. When filtering for a signal with 
a fixed amplitude, the size of $G$ is not an indicator of amplitude: it is an indicator 
only of whether we think the signal is present or absent.

This illustrates the great 
importance of having the correct information built into the signal template $h$ that 
one looks for: if one assumes a fixed given amplitude for the signal, one forces the 
statistics of the test to choose between a signal of that amplitude and nothing at all.  If 
the signal really arrives with a different amplitude, then one is using an inappropriate test 
and can get the wrong answer, particularly for the significance of the detection.  When dealing 
with astronomical sources, one does not usually know what amplitude to expect.  Such 
a search  requires a the use of a parameter for the amplitude, whose value 
has to be deduced from the data (and in particular 
from the size of $G$).  We will treat this case in \Secref{sec:parameters} below.

\subsubsection{Detection threshold}

The decision about whether the signal is present or absent is taken by looking 
at the likelihood ratio, or more conveniently its logarithm:
\begin{equation}\label{eqn:likeligauss}
\ln\Lambda(G) = \ln\fracparen{p_1(G)}{p_0(G)} = G - \frac{1}{2} d_0^2.
\end{equation}
The Neyman-Pearson test sets a suitable threshold on $\Lambda$, 
which therefore amounts to putting a threshold 
$G_0$ on the detection statistic.  Then the false alarm and detection 
probabilities are given by, respectively, 
\begin{eqnarray}
P_F & = &\frac{1}{2}\mbox{erfc}\fracparen{G_0}{\sqrt{2}d_0}
\qquad\mbox{and} \label{eqn:faprobNP}\\
P_D & = & \frac{1}{2}\left[1 - \mbox{erf}\fracparen{G_0-d_0^2}{\sqrt{2}d_0}\right], \label{eqn:detprobNP}
\end{eqnarray}
where $\mbox{erf}(x)$ is the error function and $\mbox{erfc}(x) = 1 -  \mbox{erf}(x)$ is 
its complement. (We adopt the convention that the error function is 
an odd function of its argument.)  Thus, the probabilities 
governing the detection of a known signal buried in Gaussian noise are 
completely determined by the amplitude signal-to-noise ratio $d$.

If the right-hand-side of \Eqref{eqn:likeligauss} is positive, then the likelihood ratio is 
larger than 1, and the observed data were more 
probable if the signal was present than if it was absent.  This might be a reasonable threshold 
to impose on the test: $G_0=d_0^2/2$.  
The false-alarm probability for this threshold is $0.5\;\mbox{erfc}\left(d_0/2^{3/2}\right)$, and 
the detection probability is $0.5[1-\mbox{erf}(-d_0/2^{3/2})]=0.5[1+\mbox{erf}(d_0/2^{3/2})]$.  
For an amplitude snr of $d_0=5$, we find for this threshold $P_F = 0.006$ and $P_D = 0.994$.

If one wants 
to bias the test even more against false alarms, so that one can be really confident of any detections, 
then one can choose a larger threshold.  For example, one could set $G_0=d_0^2$, 
insisting that the filter should reach the expected signal even in the presence of noise, which 
it can do, of course, only half the time.  Then 
the false-alarm probability would plummet to $0.5\;\mbox{erfc}(d_0/\sqrt{2})$, which is $3\E{-5}$ 
for our example of $d_0=5$.  But the detection probability would go down to 
$0.5[1- \mbox{erf}(0)] = 0.5$, as we expected.   

The threshold is not chosen by a magic algorithm: it must be set at a 
level that gives one (really, the scientific community generally) sufficient confidence 
that the claim of a detection is correct.  For the first gravitational wave 
detection, this threshold will probably be set fairly high.  When one 
is studying a population of sources, where one can tolerate some 
mis-identifications but one wants to get the overall population correct, 
then the threshold can be somewhat lower.  

Normally thresholds are set on $G$ at a level of several times $d_0$.  In this case, it is useful 
to know the asymptotic approximation for the complement of the error function, 
\[\mbox{erfc}(z) = 1-\mbox{erf}(z) = \frac{e^{-z^2}}{\pi\sqrt{z}}\left[1+O\left(z^{-2}\right)\right]. \]

\subsubsection{Increase of signal-to-noise ratio with time for long-duration signals}

A continuous, long-duration signal is narrow-band: it has relatively few significant Fourier 
components in its spectrum.  We saw in \Secref{sec:parseval} that the Fourier 
amplitude of such a signal increases linearly with observation time $T$.  We also 
saw that the psd $S_k$ increases linearly with $T$.  The power signal-to-noise ratio $d^2$,
therefore, in \Eqref{eqn:snrsta} also increases linearly with time.  Since we usually 
deal with amplitude signal-to-noise, it is useful to remember the rule of thumb: {\em when 
observing a narrow-band signal of long duration, the amplitude signal-to-noise ratio 
increases with the square-root of the observing time.}   If LISA can detect a binary system 
in its first year with a threshold signal-to-noise ratio of 5, then after 4 years of observing 
the signal-to-noise ratio will be 10.

The derivation we gave of the dependence of Fourier amplitudes on the length 
of the data set in \Secref{sec:parseval} was purely mathematical and in addition 
depended on the particular way we normalised the Fourier transform.  It is useful, 
therefore, to try to get a more physical feeling for this very important result.  I can 
offer two ways of looking at the calculation.  The first way is to look at the power in the 
signal, as in Parseval's theorem.  Since the signal is coherent, its Fourier amplitude 
just increases linearly with  observing time.  The noise 
behaves differently because it is incoherent.  The noise amplitude performs a random 
walk.  It accumulates only as the square-root of the time.  Therefore the amplitude 
signal-to-noise ratio increases as the ratio $T/\sqrt{T} = \sqrt{T}$.  

The second way of looking at this is to concentrate on the intrinsic signal amplitude, 
not its Fourier transform.  This does not change during the observation, so it is a 
constant.  But by performing a Fourier transform, we remove most of the noise that 
competes with the signal in the data stream by putting in Fourier bins (discrete 
Fourier coefficients) far from the signal frequency.  To be detected, the signal needs only 
compete with the noise in the frequency bin that its own frequency lies in.  Now, as 
the observing time increases, the width of this frequency bin decreases, being just 
the frequency resolution $1/T$.  It follows that the noise {\em power} in this bin also 
decreases linearly with time, since the power is uniformly distributed over the spectrum.  
Therefore, the power than the signal competes with decreases, and the r.m.s.\  amplitude 
of the noise decreases as $1/\sqrt{T}$.  This leads again to a signal-to-noise ratio that 
increases as the square-root of time.

Another way of looking at this result, which has more general applicability, is that the
snr of a signal with $n$ cycles (even if they are of uneven length, 
as in a chirping signal) is higher than the snr  
of a single cycle by roughly the factor $\sqrt{n}$.  (This assumes that the noise 
is white: clearly of some cycles are at a frequency where noise is high, then they  
should not be counted in $n$.)  Because the snr of a single 
cycle cannot be significantly enhanced by filtering, the single-cycle 
snr is just the ratio of the signal's 
intrinsic amplitude $h$ to the broad-band time-series noise of the detector.  Therefore, 
the snr of a multi-cycle 
signal is comparable to that of a single-cycle signal with 
an {\em effective amplitude} of $h_{\rm eff} = h\sqrt{n}$.  One often sees sensitivity 
diagrams for gravitational wave sources (as in, for example, \cite{thorne}) in which  
many kinds of sources are plotted on a vertical scale that is their effective amplitude.  This allows 
a comparison of signal with noise for a range of sources on a single diagram.  For 
sources that have several cycles, such a plot assumes that the matched filtering has been 
done correctly.

\subsubsection{Mismatch between template and signal}

So far we have assumed that the filter was constructed from a template $h$ that 
matched the signal exactly.  But in realistic situations this match may be 
imperfect.  Perhaps the theory of the source is incompletely understood, or 
perhaps one is using a crude representation of the signal to make the computational 
job easier.  It is easy to see what the consequence of this mismatch will be.  

The variance of the detection statistic $G$ will be the same as in the ideal case, 
\Eqref{eqn:filtervariance}, where $h$ represents the template used to construct 
the filter.  Suppose, however, that the signal arriving has a waveform $H_j$.  Then 
the actual value of $G$ will be $\sum_j H_jq_j$.  If this is close to the ideal $G$, 
then the earlier discussion will be substantially correct.  

This leads to an important observation: in the construction of a matched 
filter for a multi-cycle signal, it is more important to track the phase 
of the incoming signal than to match variations in its amplitude.  The reason 
is that the correlation sum must come out right, and this requires $q_j$ to be positive when 
the $H_j$ is positive and $q_j$ to be negative when $H_j$ is negative.  If the template and signal  
get out of phase at some time, they will make negative contributions to the total sum, and the 
value of $G$ (and hence the sensitivity of the filter) will drop quickly.   On the other hand, 
if the amplitudes are not exactly right, the power in the correlation will be affected, but only modestly.  

This becomes obvious if we think of the Fourier transform as a matched filter for 
single-frequency sinusoidal signals.  If the frequency of the template sinusoid differs from that 
of the incoming sinusoid by as little as the frequency-resolution of the observation $1/T$, 
the two waveforms will be orthogonal, and the correlation will be zero: this is mathematically the way the 
Fourier transform tells us what the true frequency is. So if the signal and template get out of phase 
by just one cycle in the observation time, the sensitivity of the test is destroyed.  This 
can be the result of a very small fractional error in the frequency of the signal.  A comparably 
small change in the amplitude will have a negligible effect on the correlation.  It is 
important to bear this lesson in mind when constructing templates for long-lived, multi-cycle signals.

\subsubsection{Noise plots}

Beginners in signal analysis, and this includes many sophisticated theoretical 
physicists, are often bewildered when they see experimentalists plot the 
noise in their gravitational wave detectors with a vertical axis labelled ``metres per 
root Hertz'' or ``$\units m\;Hz^{-1/2}$''.  What can the square-root of a Hz mean?

We are now at a place where we can understand this mystery.  These graphs 
typically plot the noise {\em amplitude} in an experiment.  Now, we have seen 
that the noise has power that is distributed smoothly over the spectrum, and 
this is measured by the psd.  We would expect, therefore, that the psd should 
have dimensions of power (say, $\units m^2$ for an output $x$ that is measured 
in metres) per unit frequency.  Our definition of the psd, $S_k$, does not have 
these units because we drop the dimension of time (or frequency) when we form 
the DFT.  The correct conversion to the actual power per unit frequency at a 
particular frequency, which we shall call $S( f )$, involves dividing $S_k$ by the 
frequency interval $\Delta  f  = 1/T$ and (for our convention for the placement of the factors of 
$N$ in the DFT) by $N^2$.  This gives [in the same sense of correspondence as in 
\Eqref{eqn:DFTandcont}]
\begin{equation}\label{eqn:continuouspsd}
S( f_k) \rightarrow \frac{T}{N^2}S_k. 
\end{equation}
This is the noise power, but it is usually more informative to plot the 
noise amplitude, in order to compare it with expected signal 
amplitudes.  The noise amplitude is the square root of $S( f )$ and 
is typically denoted by $\tilde{n}( f )$:
\begin{equation}\label{eqn:continuousnoise}
\tilde{n}( f_k) = \left[S( f_k)\right]^{(1/2)} \rightarrow \left[(T/N^2)S_k\right]^{(1/2)}.
\end{equation}
Clearly this has dimensions of amplitude (metres or whatever) per root Hz.  
The conversion to a more physical quantity, such as metres, is to multiply 
the noise amplitude by the square root of the bandwidth that the signal 
occupies, because it is in this band that the signal fights the noise, as 
we discussed in the previous section.  Then 
one gets an overall noise amplitude in metres, to compare with the signal.

\subsection{Dependence on Parameters}\label{sec:parameters}

\subsubsection{The maximum likelihood estimator}

In general we do not know everything about an expected signal: we only 
know the form of the  signal $h(t)$ as  a  function  of  
a number of parameters.  With astronomical sources, the amplitude of the signal 
and its time of arrival are almost always unknown parameters.    There may be others. 
For example, the waveform from a coalescing supermassive black-hole binary will depend 
on two further parameters: the chirp mass (a certain combination of 
the masses of the members of the binary), and the  phase  of the wave 
at the time of arrival. In such a case in order to detect the signal  
we  must also determine its parameters.  In the maximum likelihood 
method, the natural approach is to find values of the parameters of the 
family of templates that maximises the likelihood function for a given set 
of observed data.  The set of optimum parameter values is called the 
maximum likelihood estimator.

Let  ${\theta} = \{\theta_1, \theta_2, \ldots, \theta_m\}$  be the 
set of $m$ unknown parameters of the signal $h(t;{\theta})$, and 
let (as in \Eqref{eqn:detprob}) $\bf x$ stand for the observed 
data function $x_j$.  As in the case of a completely known signal  we  
consider  two  probability  density  functions $p_{\theta_0}({\bf 
x};{\theta})$ and   $p_0({\bf x})$  depending on whether a signal with parameter set $\theta_0$ 
is present or absent.  From these we form the parameter-dependent likelihood ratio 
based on a test for whether the data contains a signal with a parameter set $\theta$, that 
may or may not be the same as the parameter values $\theta_0$ that the signal really has:
\begin{equation}\label{eqn:mle}
\Lambda[{\bf x};{\theta},\theta_0]=\frac{ p_{\theta_0}({\bf x};{\theta})}{p_0({\bf x};\theta)}.
\end{equation}  
We define the {\em maximum likelihood  estimator} (MLE)   
$\hat{\theta}$  of  the  set of parameters $theta$ to be the set 
that maximises the likelihood ratio $\Lambda[{\bf x};{\theta},\theta_0]$.  
Hopefully, this will not be too different from $\theta_0$.    
The  MLE  can  be found by solving the set of simultaneous equations 
\begin{equation} 
\frac{\partial}{\partial\theta_j}\Lambda[{\bf x}; \theta_1, \ldots, 
\theta_m; \theta_0]  = 0 \qquad \mbox{for $j=1, \ldots, m$}. \label{eqn:ML} 
\end{equation}

Let us construct this function in the presence of stationary Gaussian noise.  
The appropriate filter $q_k({\theta})$ is now parameter-dependent, 
\[\tilde{q}_k(\theta) = \tilde{h}_k(\theta)/S_k.\]
This family of filters produces a family of detection statistics
\[G(\theta) = \sum_k \tilde{x}_k \tilde{q}_k^*(\theta).\]
To apply the likelihood test, let us test for the presence of 
a signal that matches our template family for some single 
set of parameter values $\theta_0$.  The pdf for the data if 
there is no signal at all is the same as we had before, \Eqref{eqn:statnone}:
\begin{eqnarray}
\mbox{absent:}\quad p_0\left(G(\theta)\right) & = & (2\pi d_\theta^{2})^{-1/2}\;\times \nonumber \\
&&\exp\left(-G(\theta)^2/2d_\theta^2\right), \label{eqn:statnoneparam} 
\end{eqnarray} 
where the variance of the random variable $G(\theta)$ is 
\begin{equation}\label{eqn:filtervarianceparam}
d_\theta^2 = \e\left[G(\theta)^2\right]\mbox{ (no signal) } = 
\sum_{k=0}^{N-1}\frac{|\tilde{h_k(\theta)}|^{2}}{S_k}.
\end{equation}

When a signal is present with parameter set $\theta_0$, the pdf depends on 
both the signal parameters and the template parameters:
\begin{eqnarray}
\mbox{present:}&\quad& p_{\theta_0}(G(\theta))  =  (2\pi 
d_\theta^{2})^{-1/2}\;\times \nonumber \\
&&\quad\exp\left[-(G(\theta)-d_\theta^{2})^2/2d_\theta^{2}\right]. \label{eqn:statsigparam} 
\end{eqnarray} 
The variance is the same as in \Eqref{eqn:filtervarianceparam}. The 
false-alarm and detection probabilities have the same form as for the 
non-parametric signal, \Eqref{eqn:faprobNP} and \Eqref{eqn:detprobNP}.

The likelihood ratio has the logarithm
\begin{eqnarray}
\ln\Lambda&=&\fracparen{p_{\theta_0}({\bf x}; \theta)}{p_0({\bf x}; \theta)} \nonumber\\
&=&G(\theta) - \frac{1}{2}d_\theta^2.\label{eqn:paramlike}
\end{eqnarray}
The MLE is the value of $\theta$ that maximises this expression for a given data set. 

Normally, one cannot maximise this as a function of a continuous set of values of 
$\theta$.  Instead, one establishes a grid of discrete values of the parameters that 
covers the important range of parameter values with sufficient density not to miss 
signals with intermediate parameter values.  If there are $m$ parameters, this 
grid is $m$-dimensional; since it may in some cases be necessary to test for hundreds 
or thousands of grid values along each dimension, the problem of finding the MLE in 
a multi-parameter template can be very demanding computationally.
 
Once the maximum value of the likelihood ratio has been found on a grid  
of filters, the spacings between parameter values may be further 
refined to get as close an approximation to the MLE as desired. 
Note that the errors in determining various parameters can be 
correlated, and this can make some parameters more difficult to determine 
than others.  These problems have been extensively discussed in the 
gravitational wave field in relation particularly to coalescing binary signals.  We 
will give references when we address this family of signals below.

\subsubsection{Amplitude as a parameter}

One parameter that is part of almost every signal LISA might search for 
is the amplitude.  Most sources could be at a range of distances, so 
the amplitude is not predictable.  Even known binary systems have 
uncertain distances, and even more uncertain inclinations, so the 
amplitude of LISA's response will not be known in advance.  Fortunately, 
the amplitude parameter is straightforward to determine.  We discuss it 
as a good concrete example of the general procedure outlined above.

Let us suppose that we are searching for a waveform described by a 
function $h(t)$, which has some suitably chosen (arbitrary) amplitude.  The 
incoming signal amplitude is $A_0$: $h(t;A_0) = A_0h(t)$.  The family of 
templates that we use has an amplitude parameter $A$: $h(t;A) = Ah(t)$.
Then the filter family will be $q(t;A)$ whose values at the sampling times are
\[\tilde{q}_k(A) = AN\tilde{h}_k/S_k.\]
Then the detection statistic is
\[G(A) = A\sum_k \tilde{x}_k\tilde{h}_k^*/S_k.\]
For this parameter, the sum has to be done only once: we can treat the 
parameter dependence analytically.  The variance of this statistic is 
\[d_A^2= A^2 d_0^2,\]
where $d_0^2$ is given in \Eqref{eqn:filtervariance}.  The standard deviation 
of the filter output is thus $A d_0$.

To deduce the snr when a signal is present, take the expectation of $G(A)$:
\[\e[G(A)] = A\sum_k \e\left[\tilde{x}_k \right] \tilde{h}_k^*/S_k.\]
Since the data stream contains a signal with amplitude $A_0$, this evaluates to
\[\e[G(A)] = AA_0d_0^2.\]
The snr is, therefore, this expectation divided by the standard deviation:
\begin{equation}\label{eqn:snramplitudeparam}
\mbox{snr} = A_0 d_0.
\end{equation}
It is satisfying that the snr depends on the intrinsic amplitude of the signal and 
not the amplitude of the parameter-dependent filter.  This is because the 
filter amplitude also appears in the noise output of the filter.  

Now we can search through the filters to determine the one that matches the 
amplitude.  This is not really necessary in this simple case, since any filter can 
be used to deduce the snr and infer whether the likelihood threshold has been 
exceeded.  But the procedure will illustrate what happens in more difficult cases.  
The logarithm of the likelihood function in this case is (from \Eqref{eqn:paramlike})
\begin{equation}\label{eqn:amplitudeloglike}
\ln\Lambda=G(A) - \frac{1}{2}A^2d_0^2.
\end{equation}
The maximum of $\Lambda$ occurs where this exponent reaches a maximum, which 
leads to the condition for the best parameter value $\hat{A}$:
\[ \sum_k \tilde{x}_k\tilde{h}_k^*/S_k - \hat{A} d_0^2 = 0.\]
This can be solved for $\hat{A}$:
\begin{equation}\label{eqn:amplitudedetemination}
\hat{A} = \frac{\sum_k \tilde{x}_k\tilde{h}_k^*/S_k}{d_0^2}.
\end{equation}

It is important to realise that the variable $\hat{A}$ is a random variable.  It has an 
expected value and a variance around that expectation.  Its expectation is similar 
to the calculation of the snr:
\begin{equation}\label{eqn:expectedamplitude}
\e\left[\hat{A}\right] = \frac{\sum_k A_0\tilde{h}_k\tilde{h}_k^*/S_k}{d_0^2} = A_0.
\end{equation}
Therefore, the most likely value we should deduce for the amplitude of the 
signal by the method of maximum likelihood is the true amplitude of the incoming signal.  

The variance of this estimate $\hat{A}$ is also important and instructive.  It is 
easy to show that 
\begin{equation}\label{eqn:varianceofamplitude}
\e\left[\hat{A}^2\right] - \left(\e\left[\hat{A}\right]\right)^2 = \frac{1}{d_0^2}.
\end{equation}
The standard deviation of the amplitude estimate is thus $1/d_0$.  To understand this, 
calculate the relative accuracy of determining the amplitude.  This is 
\[\frac{\delta\hat{A}}{\hat{A}} = \frac{1/d_0}{A_0} = \frac{1}{\mbox{snr}}.\]
This is our first example of a general rule: the accuracy with which a parameter 
can be determined improves directly with the snr with which the signal is detected.
If a signal is detected with snr of 10, then its amplitude can be determined to 
an accuracy of 10\%.

Thresholds must still be set with regard to the false-alarm probability.  For the 
statistic $G(A)$, suppose the threshold is set at $G_0(A)$.  Then, since the 
variance is $d_A^2$, the false-alarm probability is 
\[P_F(A) = \frac{1}{2}\mbox{erfc}\left(G_0(A)/\sqrt{2}d_A\right).\]
If we set the threshold to a given multiple of the standard deviation $d_A$ of the filter 
output, then this probability will be independent of which parameter value $A$ we 
have chosen, and the estimation of false-alarm probabilities will be the same 
as for the case with no parameters.

\subsubsection{Time-of-arrival as a parameter}

Most signals have structure in time that makes some times different from others. 
If a signal has a specific beginning, we call that the time-of-arrival.   The signal from the coalescence 
of two black holes does not start at a special time, but it finishes at the coalescence time.  
A signal from a chirping binary may not start or finish at any time during the observation, 
but one can assign a time to it by measuring the time when the frequency reaches some 
particular fiducial frequency.  In general, we will refer to these times as fiducial times.  But 
the general problem is more often called the determination of the time-of-arrival of 
a signal.  The fiducial time of a signal is as important a parameter as the amplitude: 
in astronomy we can almost never predict when a signal should arrive.

The family of signals that arrive at different times but are otherwise identical is a simple family, just 
obtained from one another by time-translation.  We begin with a particular filter for one of them, say 
where the fiducial time is at $t=0$.  Let us call this $q(t)$.  Then the filter for a signal that has fiducial 
time $t_c=c\Delta t$ is just $q(t-t_c)$.  
(This assumes that the noise, which weights the filter, is the same 
at all times: stationary noise again.)  Then the detection statistic 
for a signal with arbitrary fiducial time 
is (compare with \Eqref{eqn:Gdef})
\begin{equation}\label{eqn:fiducialtime}
G(t_c) =  \sum_{j} x(t_j)\,q(t_j-t_c).
\end{equation}
This is just a convolution of the data with the filter: the filter slides through the data, performing a 
sum at each step.  If the statistic $G(t_c)$ becomes exceptionally large, then we should 
look at that time for the signal.  Each value of $t_c$ represents a different filter, and the statistics 
of each one are as before. 

The MLE follows from constructing the logarithm of the likelihood function, which in this case is $\ln\Lambda = G(t_c) - d_0^2/2$.  Taking the derivative with respect to $t_c$ involves only 
differentiating the first term, since the second is constant.  Finding the fiducial time 
therefore amounts to looking for a local maximum in $t_c$ of the detection statistic.  If 
this maximum exceeds the detection threshold, one has confidence that the signal arrived 
with that fiducial time.

The appropriate threshold for  $G(t_c)$ must take into 
account the possibility that, among the $N$ random values $G(t_c)$ for the N possible times $t_c$, 
at least one will reach the observed value of $G$ even if there is no signal.  That is, 
because we perform the filtering $N$ times, the probability that at least one such filter will 
cross any given threshold even in the absence of a signal is $N$ times larger.  The threshold
must be raised to compensate for this. 

The estimation of the accuracy of the measurement of the fiducial time would occupy too 
much space here.  I refer interested students to a large literature on this and related 
subjects:  \cite{schutzblair,cutlerflanagan,vechhionicholson}.

While it might seem that filtering for $N$ different fiducial times would be a difficult task in 
a long observation, it is actually much easier than it might seem, thanks to the efficiency 
of the fast Fourier transform algorithm.  The use of this for determining fiducial times is described 
below.

\section{LOOKING FOR EXPECTED SIGNALS WITH LISA}

Now we turn to the specific problems presented by the signals we expect LISA to 
see.  Coherent signals, such as those from binaries, will be found by matched filtering.  
For these we  must examine what the filter should look like, what its parameters must 
be, and how much computing power will be needed to cover a reasonable range of parameter 
values.  It is convenient to distinguish between short-duration signals, which last for less 
than the duration of the observing period of a few years, and long-duration signals, which are 
present for the whole observation.  We must also consider incoherent signals, which are the 
gravitational wave backgrounds due perhaps to binary systems in the Galaxy and in external 
galaxies, or to gravitational waves from the Big Bang.  

\subsection{Phase and Amplitude Modulation by Detector Motion}
 
Even short-duration signals will normally last long enough for the detector to move around the Sun by 
a significant amount during the observation, introducing a Doppler shift in the observed frequency.   Because all coherent signals will be affected by the detector's motion, we begin by examining its effect.

As LISA orbits the Sun, it moves toward and away from any fixed source.  This produces a Doppler shift in the apparent frequency of signal; the shift depends on the position of the source in the sky.  Moreover, the plane containing LISA's arms rotates once per year, and within that plane the orientation of the arms rotates as well.  The effect is to scan any given source along a path through the antenna pattern of each of LISA's interferometers.   This produces a modulation of the amplitude of the response of each detector to the amplitude of the wave.  This modulation depends on the position of the source in the sky and on the polarisation of the wave.  Any other space-based gravitational wave detector will inevitably experience  analogous effects, though they may differ in detail from those of LISA.  Our discussion will also apply 
in its basics to ground-based detectors, which are carried through the Solar System by the motion of 
Earth. 

If we know, or think we know, the way the intrinsic frequency and amplitude of the detected waves depend on time, then we can use the position-depend\-ence of the frequency modulation to infer the position of the source, and we can use the polarisation-depend\-ence of the amplitude modulation to infer the intrinsic polarisation of the wave.  These effects are therefore crucial in enabling LISA to extract useful information from its observations.

The amplitude modulation depends on the position of the source on the sky, but for a long-lived 
source LISA's orbital motion performs an average over many positions, so the average amplitude of 
LISA's response to a source is only weakly dependent on the source's position.  In the sensitivity 
diagram of LISA (Figure~2 of my lecture at this school on sources for LISA; also in \cite{LISA1996}), 
the detection threshold is drawn at the intrinsic amplitude a signal would have 
to have in order that LISA's average response would be  $5\sigma$ above the LISA noise.  This level 
is actually at $5\sqrt{5}$ above the LISA raw noise curve.  The reason for this is that the vertical 
scale in this figure is the intrinsic amplitude of the signal, not the response of the detector.

The Doppler effect is more fruitfully regarded as a phase modulation, by which I mean that the detector 
moves through the highly-regular wave pattern in a way that brings it across successive peaks (phase zero) in varying amounts of time.  Using this picture, it is clear that there is significant phase modulation during a LISA observation only if the wavelength of the gravitational wave is comparable to or smaller than the diameter of LISA's orbit, $\lambda_{\rm crit} = 2\units AU$.  This means that phase modulation is unimportant at gravitational wave frequencies below about $ f_{\rm crit} = c/\lambda_{\rm crit} = 1\units mHz$.  Below this frequency, all directional information comes from amplitude modulation (\cite{peterseim}).

\subsubsection{Analogy with the interference between two signals}

Phase modulation is the basis of LISA's ability to tell the direction of a source on the sky.  To see how 
this works, it is helpful to begin with an analogy that is more familiar, but which is in essence the 
same effect: signal interference patterns.

Most interesting sources for LISA lie below about 10~mHz, for which LISA's orbit spans at most 10 gravitational wavelengths.  To see what this means for data analysis, consider two sources of gravitational waves with identical (and constant) frequencies, but in different positions on the sky.  Radiation from the two sources creates an  interference pattern in the solar system: there are places where they happen to add to each other (constructive interference) and places where they subtract (destructive interference).  

If the wavelength is short compared to 1~AU, then there will be many such maxima and minima within the orbit of LISA, provided the waves arrive from rather different directions.  If the waves arrive from almost the same direction, then the regions of constructive and destructive interference will be separated by much more than a wavelength, and LISA will not be able to tell that there are two different sources.

Now, a {\em stationary} detector would experience a coherent superposition of the two signals, which might interfere constructively or destructively, depending on the exact position of the detector.  But as the detector does not move, it will never know that there are two different sources: it will simply receive a single signal with a certain phase and amplitude.

A {\em moving} detector like LISA, by contrast, will sample the interference pattern of the radiation and see that there are two different sources.  This will only be possible if the sources are sufficiently far apart on the sky for the interference pattern to be ``bumpy'' within LISA's orbit.  This effectively means that the sources must be separated by an angle greater than the ratio $\lambda_{\rm gw}/(1\units AU)$, in radians.  

The power that a detector sees will be the power in the sum of the two signals.  Source 1, in a location on the sky that we call ${\bf \theta}_1$, produces a waveform $h_1(t; {\bf \theta}_1)$ in the detector as it moves through the wave pattern, and source 2 (at the position we call ${\bf \theta}_2$) produces $h_2(t; {\bf \theta}_2)$.  The power at time $t_j$ is (suppressing the angular arguments here)
\begin{eqnarray}
P_j & = & \left[h_1(t_j) + h_2(t_j)\right]^2  \nonumber \\
& = & \left[h_1(t_j) \right]^2  +  \left[h_2(t_j) \right]^2 +  2 h_1(t_j)h_2(t_j). \label{eqn:powerinterference}
\end{eqnarray}
The first two terms are constant on time-scales longer than the period of the wave (at most a few hours in the case of LISA), but the final term is the one that will change on the orbital time-scale because of the motion of the detector: 
\begin{equation}\label{eqn:interferenceterm}
\mbox{Interference term } = h_1(t_j; {\bf \theta}_1)h_2(t_j; {\bf \theta}_2).
\end{equation}
This is the effect of the interference.

\subsubsection{Filtering with the wrong filter: direction-finding with LISA}

Now, in reality there will not be two sources with identical frequencies.  But this thought experiment helps us to understand how LISA can use phase modulation to determine the direction to a single source.  In the data analysis, we will pass the data through a filter that represents the waveform expected from a source in a particular direction.  Let us suppose that this is in the position of source 1 above, so that we use $h_1(t; {\bf \theta}_1)$ as the filter. (For simplicity, let us assume white noise, where the filter is the expected signal.)  Now suppose there really is a gravitational wave of this frequency in the data, but that it comes from the position of source 2.  Then the data will contain the signal $h_2(t; {\bf \theta}_2)$. When we filter the data, the expectation of the output of the filter is given by \Eqref{eqn:Gdef}:
\begin{equation}\label{eqn:filterinterference}
G = \sum_j h_1(t_j; {\bf \theta}_1)h_2(t_j; {\bf \theta}_2).
\end{equation}
But this is just the interference term in \Eqref{eqn:powerinterference}.  Therefore, the problem of finding a signal with a filter that assumes the wrong angular position is just the same as seeing the interference pattern between the two signals.  This means that we will have a good response if the position assumed by the filter and the position of the true signal are close enough on the sky.  If they are not, then there will be a poor response.  This effect gives LISA its directional capabilities, but at the price of having to perform the filtering for many independent locations on the sky.

This means that the angular resolution of LISA when looking at a weak source, near its noise level, will be at best of order 1/10 radians.  However, this resolution improves 
essentially directly with improving snr, because for a strong signal one can notice small variations in the output of the filter when its position is slightly changed.  Therefore, for a coalescing black hole binary with a signal-to-noise ratio of $10^4$, the angular resolution could  in principle be as good as $10^{-5}$~rad, or a few arcseconds.  This is, however, unrealistic, for three reasons: first, such sources spend most of their time (and their accumulated signal) at sub-mHz frequencies, where there is no detectable phase modulation; second, there is a correlation of errors between position variables and other parameters that degrades the accuracy; and third, noise from galactic binaries may degrade both the overall signal-to-noise ratio and the accuracy of position-sensing.  Realistic estimates (\cite{cutler}) give typical angular accuracies of about 1 degree for massive black-hole coalescences.  For signals 
at a constant frequency, like binaries, the situation can be much better, with accuracies approaching 
the arcminute scale.

\subsection{Analysis of Data for Short-Duration Signals}

Some of LISA's most exciting signals, those from coalescing black-hole binaries, will have lifetimes 
shorter than the duration of the experiment if they enter the LISA frequency window at all.  Therefore 
one of the most elementary parameters associated with the signal is the time of coalescence, $t_c$.  
This is a signal that requires a fiducial-time parameter, as discussed earlier.  

\subsubsection{Construction of coalescing binary templates}

The merger of two black holes is a very complicated dynamical event, and one might worry that 
it will not be possible to pick these events out effectively because our filters would be crude.  This 
is not the case, however, because the greatest part of the snr of such an event comes from the 
gravitational radiation emitted by the decaying orbit, before the merger event.  The estimates of 
LISA's sensitivity to mergers have been made assuming that all the snr comes from the orbit.  
Any further radiation from the merger event, which is really the goal of the observation, will 
be a bonus insofar as it contributes to the detectability of the event.  

But with such good 
snr from the orbital radiation, it will be possible to predict accurately the fiducial time, {\em i.e.} 
the moment when we expect to receive the 
radiation from the merger of the two black holes.  Even if that radiation is relatively 
weak,  we may then be able to study it in detail.  By the time that LISA flies, one hopes 
that the numerical simulation of black hole mergers now being developed to run on 
supercomputers will produce accurate templates, so that we can study the details of 
the merger as a test of black-hole theory in general relativity.  

Cutler (private communication) 
has pointed out another bonus of having a good coalescence time.  If the data analysis 
can be done fast enough to predict the coalescence before it occurs, then astronomers 
can be alerted to the imminent event and  given a crude position.  It may well be that 
the merger will be accompanied by some kind of optical or X-ray display.

The construction of templates even for the orbital radiation is itself a difficult task.  The orbital 
system is approximated as two point masses orbiting in general relativity.  The holes may make 
several hundred orbits while LISA watches them, before they merge. During this time, 
the template must not get out of phase with the orbit, so the orbital decay must be tracked 
accurately.  The simple Newtonian and quadrupole approximations that suffice to 
give factor-of-two accuracy for most sources (see my first lecture at this school) do not give the 
required accuracy here.  An approximation scheme called the post-Newtonian method must 
be used, and it must be pushed to very high order to get good results.  This is being done 
now in order to provide good templates for ground-based observations of neutron-star binaries, 
and the same work will be applicable to LISA's sources.  See \cite{3minutes,blanchet,Blanchet} for 
more details.

As remarked earlier, the job of filtering for each different coalescence time is made much 
more tractable by using the Fast Fourier Transform algorithm.  This is so important that it 
must be described here.

\subsubsection{Filtering using the Fast Fourier Transform: benefits and pitfalls}

The assumption of stationary noise leads us naturally to use DFT's for our  
analysis, since (as we have seen) the noise at different 
frequencies is uncorrelated.  But there is an additional reason for using 
DFT's, which is that the convolutions needed for filtering for different fiducial times 
in \Eqref{eqn:fiducialtime} are most rapidly done using the well-known 
Fast Fourier Transform (FFT) algorithm.  

An expression like the standard DFT,
\[\tilde{h}_k=\sum_{j=0}^{N-1}h_je^{-\twopii jk/N},\]
would require $N$ multiplications and $N$ additions for the sum that produces each 
element of the transform.  Since there are $N$ values of $k$ for which this needs to 
be done, there are of order $N^2$ operations to compute the DFT in a straightforward 
manner.  The FFT algorithm, by making use of symmetries of the unit complex numbers 
$\exp(-\twopii jk/N)$, reduces this to a number of order $N\log_2N$.  For a data set 
containing $N=10^8$ data values, as LISA will have (see below), this is a speedup of 
a factor of more than one million.  

The same is true of the convolution and correlation.  If we want to evaluate a filter with a variable 
fiducial time, 
\[ G(t_c) =  \sum_{j} x(t_j)\,q(t_j-t_c),\]
then we must again perform $N$ operations for each value of $t_c$, and there will be 
$N$ possible values of $t_c$.  But if we use the correlation theorem, \Eqref{eqn:discretecorrthm1}, 
we have only one multiplication for each value of the Fourier transform of the correlation, 
\[\tilde{G}_k = \tilde{x}_k\tilde{q}_k^*.\]
In addition to this we have to perform three Fourier transforms: two to obtain $\tilde{x}$ and $\tilde{q}$ 
(although the latter may already be available from \Eqref{eqn:filtersolve}), and a third to go back 
from $\tilde{G}$ to $G$.  Adding these up still gives an operation count of order $N\log_2N$ rather 
than $N^2$.  

So matched filtering for signals with unknown fiducial times is usually done with 
FFT's.  However, there is a subtle problem that the user must beware of.  This is the problem of 
wraparound, or end effects in the correlation.  The correlation of the data with the moving filter, 
\[G(t_c) =  \sum_{j} x(t_j)\,q(t_j-t_c),\]
is only well-defined if $t_j-t_c$ is within the range of times defined for the filter $q$, which cannot be 
larger than $N$.  Therefore, for any $t_c$ there will be a range of values of $j$ for which $q(t_j-t_c)$ is not defined.  When the filter $q$ is of short duration compared to the length of the 
data set, the values of $q$ will be zero at almost all these points, so the correlation can be 
performed without error.  But there will always be some values of $t_c$, usually at the beginning 
or the end of the range of times, where the summation reaches the end of the range for $t_j$ without 
reaching the end of the duration of the filter $q$.  In this case, the correlation is not well-defined.  In 
fact, the Fourier method of calculating the correlation makes everything periodic outside the range of 
data values, so it wraps the filter around to the other end and convolves the remaining non-zero part  
of the filter with data from the other end of the range of observation.  These will clearly not be correct 
values and must either be ignored or corrected in some way.

\subsubsection{Computational demands of filtering}

Suppose there is a single family of waveforms that one wants to look for, and suppose we have 
3 years of LISA data to look in.  The data need only be sampled effectively at about 1~Hz, since 
the upper limit on the LISA band is well below 1~Hz.  So in 3 years there will be $10^8$ data 
samples.  To search the whole data set for a single signal of unknown fiducial time would 
require of order $10^{10}$ multiplications.  On a modern high-performance workstation (not even 
a parallel supercomputer), with a memory of 1~GB and 
a speed of 1~Gflop, this would take only 10 seconds.  On computers that will be cheap 
by the time LISA flies, this will be done in a small fraction of a second.

Of course, there will be many different waveforms, not just one.  Coalescing binary black holes 
could be visible to LISA over a range of chirp masses (perhaps requiring filtering for several 
thousand different chirp mass values), over the whole sky (perhaps $10^3$ separate patches of 
0.1~rad on a side), and further parameters to describe their spins and so on.  Even so, a 
single workstation today could do the entire data set in a time comparable to the time 
it took to acquire the data.

But coalescing binaries are not as big a family as the family of waveforms that might be produced 
by compact masses falling into massive black holes.  Here the orbits will not be circular, and 
the interaction between the orbit of the compact object and the spin of the hole will lead to  very 
complex waveforms extending for thousands of observable orbits.  
So far we do not have very reliable estimates of the size of the parameter 
space that will need to be searched, but it is likely to be enormously larger than that for 
black-hole coalescences.  Even so, assuming a doubling of computer speed every $1.5$~years 
between now and, say, 2010 (to pick an optimistic time for data analysis to begin), we could 
afford the parameter space to be 1000 times larger and we could still do the search with 4 workstations.  

A more serious, and so far unexplored, problem associated with searching for compact masses may be 
the nearly-chaotic nature of their orbits around a rapidly spinning black hole.  A small change in the 
initial data for the orbit may make a very large change in the nature of the 
orbit --- particularly its plane --- many orbital periods 
later.  The key requirement of matched filtering, that it must keep in phase with the real orbit,  
may limit filters to being built out of pieces that are only a few periods long, and which 
must be joined together in unpredictable ways.  It may be hard to fit real signals accurately in this way.

\subsection{Looking for Galactic Binaries}

For a long-lived signal like that expected from a binary system, there is no real fiducial 
time, and the problem of sliding a filter through the data does not arise.  Instead, we 
would just expect to do a single Fourier transform of the data and look for a peak 
at a fixed frequency, correcting perhaps for a small chirp (change in the frequency 
during the observation).  

This is too simplistic, however, because of the induced phase modulation that 
we discussed earlier.  The simplest way to correct for this would be to try to 
reconstruct the data that a detector at rest in a single location (such as the 
barycentre of the Solar System) would observe.  This can be done from the 
data registered by the real, moving detector only if we assume a particular 
location on the sky for the source.  As discussed above, this must 
be done independently for something like $10^3$ different locations.  Once 
a signal has been identified, its position can be refined to a smaller area, 
inversely proportional to the square of the snr, but for the initial search 
the sky can be divided into about 1000 patches.

Once the reconstructed data set is obtained, then an FFT will reveal any 
constant-frequency binaries.  Any chirping binaries would have to be 
looked for with other techniques, either matched filtering or time-domain 
resampling.  Even so, the amount of numerical work involved would be 
much less than that involved in searching for compact objects falling 
into massive black holes. 

This is a great contrast to the situation for ground-based detectors.  Because 
they operate at a higher frequency, there are many more independent locations 
on the sky, and the parameter space that needs to be searched for 
continuous sources is enormous (\cite{schutzblair,patrick}).

For the purposes of 
predicting what LISA can see, the LISA team has adopted a threshold of 
$5\sigma$ for constant-frequency sources.  If the noise amplitude is Gaussian, then 
the pdf of the power spectrum is a simple exponential (since it is the sum of the squares 
of two Gaussian-distributed Fourier amplitudes).  If we set a threshold of 
5 on the spectral amplitude (which is what is plotted in the LISA sensitivity 
diagram), then this is a threshold of 25 on the power spectrum.  The false-alarm 
probability associated with this threshold is 
of order $10^{-11}$.  Since the frequency resolution in a 1-year observation 
is $3\E{-8}\units Hz$, and the LISA bandwidth is no larger than 0.1~Hz, there 
are no more than $3\E{6}$ independent elements in the DFT.  The probability that at least 
one frequency component will reach  $5\sigma$ is therefore less than $10^{-4}$.
This seems like a safe threshold to set.

As part of its observation of a binary system, LISA will determine the polarisation of the 
signal.  This will tell us the angle of inclination of the binary orbit to the line of sight 
to the system.  This is an important parameter for modelling binaries, and one of 
the hardest to obtain by optical observations.  Unless the system undergoes 
eclipses, radial-velocity measurements of the orbits of binary stars cannot 
distinguish between  a relatively edge-on system or one seen from near its 
pole.  If the latter orientation were correct, then the true velocities in the 
system would be much larger than the projected radial velocities observed from 
a spectrum, and the masses of the stars would have to be correspondingly higher.  
By contributing the actual angle of inclination, LISA will help astronomers pin 
down the masses of many known binaries, as well as of binaries that LISA 
identifies for the first time.   The resulting information should have many implications 
for improving our understanding of binary evolution, and this will lead to improved 
predictions of black-hole formation rates, supernova rates, and so on.

\subsection{Detecting a Stochastic Background} 

LISA  will find a random background of gravitational waves if the noise in the 
waves exceeds the noise in the instrument.  Unlike ground-based detectors, 
which will use cross-correlation techniques to dig below instrumental noise, 
LISA is an isolated detector and can only see noise sources above its 
intrinsic noise.  (There are two LISA interferometers, but correlating them has limited 
effectiveness: see the next section.)  

For this to work, one must have confidence that one understands the instrumental 
noise.  This can be tested by various consistency checks on the spacecraft, but 
in the end it is largely a matter of confidence in our understanding of the instrument.  

If the background is due to binary star systems, there are additional consistency 
checks one can make.  First, the amplitude of noise should vary with time as LISA 
turns in its orbit and presents different parts of its antenna pattern to the Galaxy.  Moreover, 
this is a confusion-limited background consisting of many individual stars.  Some 
members of this population will be unusually close to LISA and/or at an unusually 
high frequency, so that they will stand out from the background and be studied 
individually.  From the statistics of the individual binaries it will be possible to 
infer where the confusion background will be found, and thereby to identify it as 
gravitational wave noise rather than instrumental noise.  But if the stochastic 
waves are cosmological and isotropic, then we have no such consistency 
checks in the data.

\subsection{Using Both LISA Interferometers}

Ground-based gravitational wave detectors rely on independent observations of events 
by two or more detectors (``coincidences'') to gain confidence in the detection and to increase the 
information they obtain from an event. LISA can similarly gain from using data from both its 
interferometers.  

There is are two important differences between LISA and ground-based interferometer networks.  First, because LISA's two interferometers share a common arm, one cannot assume the noise in them is 
independent.  Therefore, when working close to the detection limit, one should not expect to gain much confidence in seeing an event in both data streams.  (Of course, if an event is seen in only one stream, 
and it lasts long enough for LISA to change its orientation and hence its polarisation, then one can 
probably assume the event was a noise event.)  Second, LISA's interferometers are in the same location, 
so there is no time-difference between signals in the two instruments that could be used (as on the 
ground) for direction-finding.

Despite these differences, there is usefully different information in the two data streams.  From the 
point of view of their gravitational wave responses, the two instruments are independent.  That is not to 
say that they are orthogonal in some sense, but simply that any given gravitational wave arriving along a direction perpendicular to the plane of LISA can be expressed as a linear combination of the two detector responses. As LISA rotates, therefore, it can use the two different responses to define the polarisation 
and to look for changes in the intrinsic polarisation of the signal.

In practice, despite the fact that a single LISA instrument can sense polarisation (by watching the amplitude  change as the detector rotates), the contribution from the second 
instrument will be very important.  In their study of signals from coalescing black holes, \cite{cutler} found that the second instrument helped to make a much cleaner untangling of the various parameters, so that the accuracy of determining the distance to the source improved 
enormously.    

I believe that the second detector is likely to be very important in 
detecting waves from compact objects spiralling into 
massive black holes.  Because of the coupling of the orbital angular momentum to 
the spin of the hole, the plane of the orbit can change dramatically during the event, and this will 
change the polarisation of the wave.  This can happen on timescales much less than the orbital 
timescale of LISA, so that LISA cannot use its changing orientation to sense polarisation.  Once the 
source's direction has been identified using phase modulation, then the polarisation will be 
determined by the two detectors operating together.  This will make the determination of other 
parameters much cleaner and easier: the
intrinsic amplitude of the signal, the distance to the source, and the direction.  
Then we can look for an association with a galaxy 
or cluster of galaxies.

When sensing a stochastic background, two interferometers could in principle do much better 
than one: by cross-correlating the noise in the two detectors, one eliminates independent noise
sources and finds only the correlated noise, which could come from gravitational waves that 
stimulate both detectors.  This is how ground-based detectors will search for a background 
at amplitudes far below their individual noise levels (\cite{flanagan}).  
But this does not work for LISA: sharing a 
common arm, the two interferometers will automatically have correlated instrumental noise.  
Coincidentally, for the $60^\circ$ angle between LISA's arms, the signal-to-noise ratio after 
correlation of the two instruments will be the same as in each individual instrument.  So if 
a gravitational wave background is hidden in an individual data stream by the instrumental noise, 
and if that noise is at the same level in all arms, then the background will be just as hidden 
in the correlation of the two outputs. 

However, the two interferometers can in principle help to determine 
whether the background is isotropic or not, {\em i.e.} to distinguish  between a cosmological 
background (isotropic) and a background due to binary systems (stronger in directions toward the 
plane of the Galaxy).  As LISA moves, it presents a different part of the antenna pattern to the 
Galaxy for one detector than for the other.  If the detector noise is dominated by the 
Galactic background, then the noise should go up and down in the right way.  The second 
detector would provide confirmation of this if its noise went up and down as well, but 
with a different phase.

\section{Conclusions}

I have introduced the elements of signal analysis as they are used for gravitational wave data from 
interferometers.  The basic techniques are the same for ground-based detectors as for LISA, 
but the data set of LISA will be much smaller (because of the lower observing frequency) 
and the data analysis problem will not be so demanding.  However, there are some 
problems, as yet unsolved, that are unique to LISA.  Most difficult is the question of how 
to devise suitable filters for the signals from compact stars falling into massive black holes 
in the centres of distant galaxies.  This is one of the most likely sources for LISA as well 
as one of the most important in terms of the fundamental information that the signals 
contain.  It will be important, as LISA develops, to find a suitable filtering method that 
does not lose many of these events.

There is one kind of source I have not discussed here, and that is the one we don't expect.  
Given the kinds of sources that inhabit the LISA frequency band, any exotic 
process that produces signals we did not expect will not only be interesting: it might 
be revolutionary.  It will therefore be an important part of the development of a data analysis system 
for LISA to build into it the capability of responding to unexpected events.  Here the notion 
of an event must be vague, so the event must stand up above noise in an unmistakable way.  
But the noise need not be the raw time-series noise: general families of filters, such as 
time-frequency methods (chirp filters, wavelets, bispectrum), nonlinear adaptive methods, and 
other general approaches can well be used to identify unexpected events.  The important 
consideration here are that these methods must be defined ahead of time, and their 
statistics must be understood.  Then it will be possible to claim that a statistically 
significant event was real.

It is possible that by the time LISA flies, gravitational waves will have been detected by ground-based 
detectors.  If this is not the case, then LISA needs to make reliable detection a high
priority, selecting filtering methods and thresholds to minimize the chance of a false alarm. The same 
will be true when looking for rare but strong events, such as black-hole coalescences in 
the centres of distant galaxies.   When 
LISA turns to the study of sources rather their mere detection, and is dealing with numerous 
populations like the galactic binaries and the compact objects falling into massive black holes, 
then the criteria become a little different, and LISA can begin observing closer to its detection 
threshold.  In the case of the galactic binaries that generate confusion noise, LISA will go below the 
conventional detection threshold in order to see how the nearby members of this class blend 
into the confused background.  

LISA's data analysis will probably  not be computationally demanding, but it will nevertheless 
require care and respect for the statistics of detection and estimation.  LISA is not likely to be 
followed soon by another mission, so its observations will be the only time the low-frequency 
gravitational wave window is opened for some time. The task of the data analysis team 
will be to give us the sharpest possible vision in this window.  What we see through that window has 
the possibility of fundamentally changing our view of Nature.  

\section*{ACKNOWLEDGMENTS}

It is a pleasure to acknowledge helpful comments on an earlier draft of this paper by Pia Astone, Peter 
Bender, Andrzej Krolak, and Maria Alessandra Papa.


\begin{thebibliography}{}

\newcommand{\MARCG}{Marck, J.-A., Lasota, J.-P., eds., {\em Relativistic Gravitation and  Gravitational Radiation}, (Cambridge University Press, 1997)}

\bibitem[\protect\astroncite{Bender et~al.}{1996}]{LISA1996} Bender P, Ciufolini I, Danzmann K, Folkner W M, Hough J, Robertson D, R\"udiger A, Sandford M C W, Schilling R, Schutz B F, Stebbins R, Sumner T, Touboul P, Vitale S, Ward H, and Winkler W,  {\em LISA: Pre-Phase A Report (MPQ 208)} (Max-Planck-Institut f\"ur Quantenoptik, Garching, Germany, 1996).

\bibitem[\protect\astroncite{Blanchet et~al.}{1995}]{blanchet}Blanchet, L., Damour, T., Iyer, B. R., Will, C. M., Wiseman, A. G., ``Gravitational-Radiation Damping of Compact Binary Systems to Second Post-Newtonian Order'', {\em Phys. Rev. Lett.}, {\bf 74},  3515  (1995).

\bibitem[\protect\astroncite{Blanchet}{1997}]{Blanchet}Blanchet, L., ``Gravitational radiation from relativistic sources'', in \MARCG.

\bibitem[\protect\astroncite{Bracewell}{1978}]{bracewell} Bracewell, R. N., {\em The Fourier Transform and its Applications (2nd edition)}, (McGraw-Hill, New York, 1978).

\bibitem[\protect\astroncite{Brady et~al.}{1997}]{patrick} Brady, P.R., Creighton, T., Cutler, C., and  Schutz, B.F., ``Searching for periodic sources with LIGO'', submitted to {\em Phys Rev D} (1997).

\bibitem[\protect\astroncite{Cutler et~al.}{1993}]{3minutes}Cutler, C., Apostolatos, T.A, Bildsten, L., Finn, L.S., Flanagan, E.E., Kennefick, D., Markovic, D.M., Ori, A., Poisson, E., Sussman, G.J., Thorne, K.S., ``The Last Three Minutes: Issues in Gravitational Wave Measurements of Coalescing Binaries'', {\em Phys. Rev. Lett.}, {\bf 70},  2984--2987  (1993).

\bibitem[\protect\astroncite{Cutler \& Flanagan}{1995}]{cutlerflanagan}Cutler, C., Flanagan, E.E.,  {\em Phys. Rev.}, {\bf D49},  2658  (1995).

\bibitem[\protect\astroncite{Cutler \& Vecchio}{1997}]{cutler} Cutler, C., and Vecchio, A., in preparation (1997).

\bibitem[\protect\astroncite{Davis}{1989}]{Davis1989} Davis, M.\ H.\ A.\ (1989), in {\em Gravitational Wave Data Analysis}, ed. B.F.\ Schutz (Kluwer Academic Publishers, Dordrecht) , pp.\ 73--94.

\bibitem[\protect\astroncite{Dickson \& Schutz}{1995}]{dickson} Dickson, C.A., and Schutz, B.F., ``A reassessment of the reported correlations between gravitational waves and neutrinos associated with SN1987A'', {\em Phys. Rev. D}, {\bf 51}, 2644--2668 (1995).

\bibitem[\protect\astroncite{Flanagan}{1993}]{flanagan}Flanagan, E.E., ``Sensitivity of the Laser Interferometer Gravitational-Wave Observatory to a stochastic background, and its dependence on the detector orientations'', {\em Phys. Rev. D.}, {\bf 48}, 2389--2407, (1993). 

\bibitem[\protect\astroncite{Helstrom}{1968}]{Helstrom1968}Helstrom,  C.\ W.\  (1968), {\em Statistical  Theory  of   Signal   Detection}, London:  Pergamon Press.

\bibitem[\protect\astroncite{Nicholson \& Vecchio}{1997}]{vechhionicholson}Nicholson, D., and Vecchio, A., ``Bayesian bounds on parameter extraction accuracy for inspiralling binary gravitational waves'', submitted (1997).

\bibitem[\protect\astroncite{Peterseim et~al.}{1997}]{peterseim} Peterseim, M., Jennrich, O., Danzmann, K., and Schutz, B.F., ``Angular resolution of LISA'', {\em Class.\ Q.\ Grav.,} {\bf 14}, 1507--1512 (1997).

\bibitem[\protect\astroncite{Schutz}{1991}]{schutzblair}Schutz, B.\ F., (1991), in Blair, D.G., ed., {\em The Detection of Gravitational Waves}, (Cambridge University Press, Cambridge England, 1991), pp. 406--452.

\bibitem[\protect\astroncite{Schutz}{1997}]{schutzleshouches} Schutz, B.F.,  ``Detection of gravitational waves'', in \MARCG.

\newcommand{\HAWKINH}{Hawking, S.W., Israel, W., eds., {\em 300 Years of Gravitation},  (Cambridge University Press, Cambridge, 1987)}                               

\bibitem[\protect\astroncite{Thorne}{1987}]{thorne} Thorne, K.S., ``Gravitational Radiation'', in \HAWKINH, p. 330--458.  

\bibitem[\protect\astroncite{Van Trees}{1968}]{VanTrees1968}Van Trees, H.\ L.\  (1968), {\em Detection Estimation and  Modulation  Theory, Part I}, New York: John Wiley and Sons, Inc.

\bibitem[\protect\astroncite{Wainstein \& Zubakhov}{1962}]{WainsteinZubakhov1962}Wainstein, L.\ A.\,  Zubakhov, V.\ B.\  (1962), {\em Extraction of Signals  from Noise}, London: Prentice Hall.

\bibitem[\protect\astroncite{Whalen}{1971}]{Whalen1971}Whalen, A.\ D.\ (1971), {\em Detection of Signals in Noise},  New York: Academic  Press.

\end{thebibliography}
\end{document}